\documentclass[conference]{IEEEtran}

\IEEEoverridecommandlockouts

\def\BibTeX{{\rm B\kern-.05em{\sc i\kern-.025em b}\kern-.08em
    T\kern-.1667em\lower.7ex\hbox{E}\kern-.125emX}}

\usepackage[utf8]{inputenc}
\usepackage{amsmath}
\usepackage{amssymb}
\usepackage{amsthm}
\usepackage{mathtools}
\usepackage{enumerate}
\usepackage{bbm}
\usepackage{algorithm}
\usepackage{algorithmic}
\usepackage{epsfig}
\usepackage{xcolor}
\usepackage{graphicx}
\usepackage{caption}
\usepackage{subcaption}
\usepackage{cases}
\usepackage{url}
\usepackage{cite}
\usepackage{mleftright}
\usepackage{xparse}
\usepackage{multirow}
\usepackage{xr}
\usepackage{comment}
\usepackage{float} 
\usepackage{multicol}
\usepackage[font=small]{caption}
\usepackage{balance}
\usepackage{enumitem}

\usepackage{soul}

\newtheorem{lemma}{\bf Lemma}[section]

\newtheorem{proposition}{\bf Proposition}[section]

\usepackage{etoolbox}

\begin{document}

\title{Exploring Performance Tradeoffs in Age-Aware Remote Monitoring with Satellites}

\author{Sunjung~Kang,
        Vishrant Tripathi,
        and Christopher G. Brinton
    \thanks{S. Kang, V. Tripathi and C. G. Brinton are with Elmore Family School of Electrical and Computer Engineering, Purdue University, IN 47907, USA.
    E-mail: \{kang392,tripathv,cgb\}@purdue.edu
        }
    }




\maketitle


\begin{abstract}
    We investigate a remote monitoring framework with multiple sensing modalities including IoT sensors on the ground, mobile UAVs in the air, and a periodically available satellite constellation. While the IoT sensors cover small areas and remain fixed, the UAVs can move between locations and cover larger areas, and the satellites can observe the entire region but have high latency and low reliability.
    We divide the deployment region into cells and model it as a graph, with the nodes representing individual cells and edges representing possible UAV mobility patterns.
    To evaluate the freshness of collected information from this graph, we adopt the Age of Information (AoI) metric, measured separately for each cell. Under a given deployment of IoT nodes and UAV mobility patterns, our objective is to ascertain whether the system should actually utilize monitoring updates from satellites -- a seemingly simple yet surprisingly elusive question. For stationary randomized scheduling policies, we develop closed-form expressions and lower bounds for the weighted-sum AoI and utilize this analysis to explore performance tradeoffs as system parameters vary. We also provide a Lyapunov style max-weight policy and detailed simulations that provide crucial insights for deploying such systems in practice.
\end{abstract}

\section{Introduction}

Remote monitoring of large geographic regions plays a key role in various application domains such as wildfire detection~\cite{meimetis2024architecture}, disaster management~\cite{higuchi2021toward}, and agricultural monitoring~\cite{yang2024integrated}. These environments often rely on heterogeneous sensing modalities such as ground-deployed Internet of Things (IoT) sensors, unmanned aerial vehicles (UAVs), and satellite-based remote sensing to collectively capture dynamic environmental information. For example, UAVs, with their high mobility and flexible deployment capabilities, can be rapidly dispatched to collect high-resolution imagery over targeted areas in precision farming. In contrast, satellites provide periodic but wide-area coverage, which makes them suitable for assessing large-scale phenomena such as flood damage after natural disasters. In wildlife monitoring, IoT sensor networks are commonly installed in forests or conservation areas, while UAVs and satellites extend the observation range by providing aerial and global perspectives, respectively.  Together, this multi-modal architecture provides complementary strengths in spatial reach, data resolution and update frequency.

However, effective resource allocation in such heterogeneous networks remains a significant challenge. Each sensing modality differs in spatial reach, size of updates generated, characteristics of wireless links, and availability. For instance, while ground sensors may offer fine-grained, real-time data, they typically suffer from limited coverage. UAVs can be dispatched to targeted areas, but have operational time and power constraints, as well as limited throughput. Satellites can capture large-scale updates, but are often available only intermittently due to orbital dynamics, and their communication channels are more prone to delays and reliability constraints~\cite{ma2023network,su2025skyoctopus,xu2024delay}. This heterogeneity introduces complex tradeoffs in selecting which data source should be used at a given time to keep the overall information state timely and accurate.

Handling all of these practical issues within one unified framework is inherently challenging, as it requires accounting for differences in sensing capabilities, availability patterns, and communication constraints across modalities. Beyond transmission scheduling, the mobility of UAVs needs to be carefully planned. In this work, we consider a simplified yet insightful model that incorporates these practical constraints and aims to address a key design question:
\begin{itemize}[topsep=2pt]
    \item[ ] \textbf{\textit{Under what conditions should transmission opportunities be allocated to satellite-based updates, alongside terrestrial sensing, to improve information freshness?}}
\end{itemize}
The answer to this question depends on multiple factors including the relative availability of satellite coverage, the reliability of the satellite channel, and the number of transmission attempts (or packets) required to successfully deliver an update compared to IoT or UAV-based sources.

A crucial component of this tradeoff is the spatial density of IoT and UAV nodes. If such nodes are sparsely deployed across a large area, relying solely on local sensing could result in long delays or information gaps. In contrast, if the satellite can transmit high-coverage updates with reasonable frequency and reliability, its use becomes more favorable. Conversely, in scenarios where dense deployments of ground and aerial sensors exist, and communication reliability is high, it may be more efficient to avoid satellite usage altogether. To explore these tradeoffs, we analyze stationary randomized policies that determine which sensing modality to activate based on system parameters. By comparing performance with and without satellite updates, our analysis identifies decision boundaries that indicate whether incorporating satellite constellations leads to improved performance or offers no additional benefit.

\subsection{Related works}

The Age of Information (AoI) has been widely studied as a measure of information freshness at the receiver side, quantifying the time elapsed since the most recent update was generated~\cite{kaul2011minimizing,guo2022age,choudhury2021aoi,hu2020aoi,sun2021aoi,tripathi2023wiswarm,yu2022age,abd2019role,corneo2019age,ngo2024timeliness}. In recent years, AoI has been increasingly used to evaluate information timeliness in time-sensitive systems, including vehicular networks~\cite{kaul2011minimizing,guo2022age}, where up-to-date status is essential, and in remote monitoring applications such as UAV-assisted sensing~\cite{choudhury2021aoi,hu2020aoi,sun2021aoi,tripathi2023wiswarm,zhen2020energy} and large-scale IoT deployments~\cite{yu2022age,abd2019role,corneo2019age,li2022age,rao2025minimum,yang2025learning}, where AoI-aware strategies help balance update frequency against resource constraints. 

\begin{figure}
    \centering
    \includegraphics[width=0.95\linewidth]{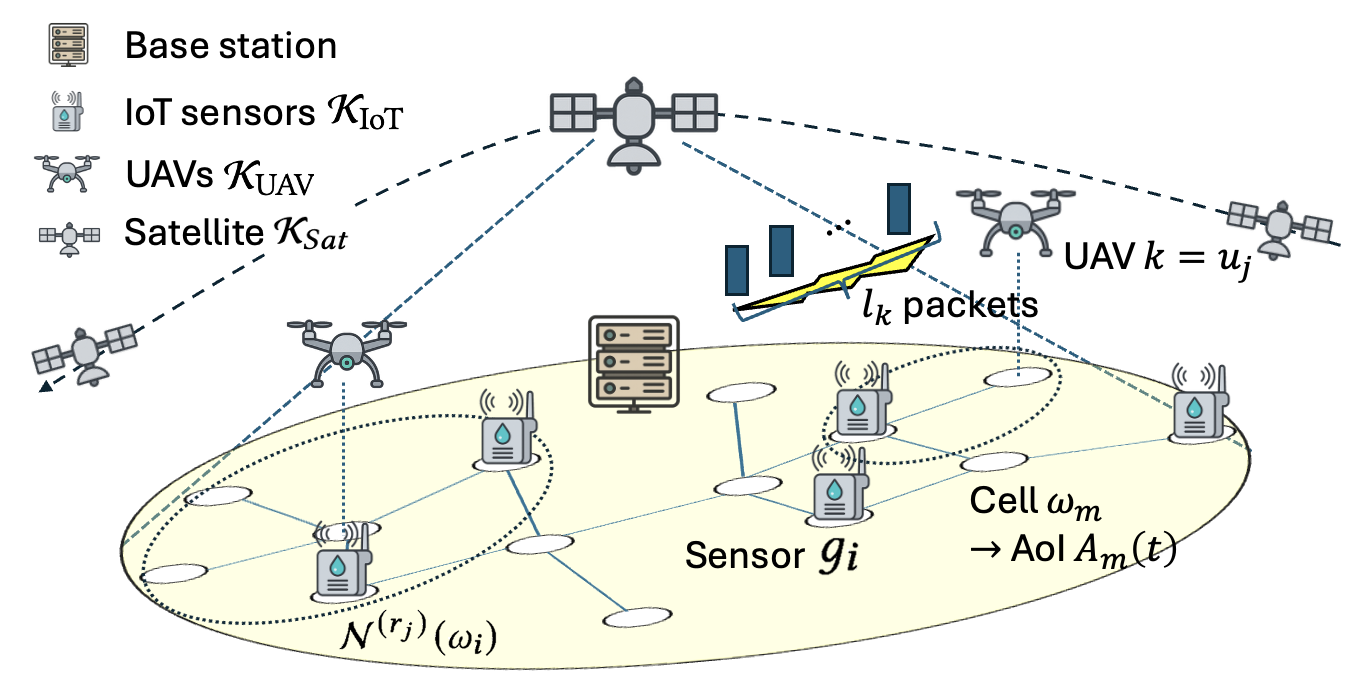}
    \caption{System model for satellite-assisted remote monitoring we consider in this paper, consisting of ground IoT sensors, mobile UAVs, and a satellite constellation. \vspace{-0.2in}}
    \label{fig:system_model}
\end{figure}

Prior works on AoI in wireless communication systems have provided the foundation for understanding how network dynamics impact information timeliness. These include studies on broadcast scheduling~\cite{kadota2018scheduling}, contention-based access in CSMA networks~\cite{maatouk2019minimizing}, and remote estimation of Markov sources~\cite{kam2018towards}. While these models offer valuable insights into queueing behavior and transmission coordination, they typically assume static sources and overlook spatial and mobility-related considerations. In contrast, more recent efforts have explored UAV-assisted sensing systems by utilizing UAV mobility to improve information freshness through trajectory optimization and update scheduling~\cite{choudhury2021aoi,hu2020aoi,sun2021aoi,tripathi2023wiswarm}. However, these UAV-based models often assume uniform packet sizes and non-overlapping sensing regions, which ignores spatial redundancy and content heterogeneity factors that are common in real-world monitoring scenarios.

To address content heterogeneity, several AoI studies have considered the effect of heterogeneous packet sizes as acknowledging that updates can differ significantly in size due to varying sensing requirements or data types. One line of work examines AoI minimization in IoT systems with non-uniform packet sizes and proposes scheduling strategies that account for differences in transmission duration~\cite{zhou2019minimizing}. Related efforts explore update scheduling under general size distributions and dynamic channel conditions~\cite{li2019minimizing,li2021minimizing}. Other works have developed low-complexity scheduling algorithms for networks with reliable channels and non-uniform update lengths, including Whittle index-based allocation strategies and structure-aware algorithms using value iteration~\cite{tripathi2021computation}. While these approaches offer useful structural insights and algorithmic frameworks, they often assume that complete updates must be transmitted without interruption. Moreover, they do not consider spatial redundancy, correlated sources, or heterogeneous availability across sensing modalities.

Another line of work considers AoI scheduling for correlated sources, where updates from multiple sensors may carry redundant information due to overlapping sensing regions. This setting is particularly relevant in remote monitoring systems where multiple devices observe the same or partially overlapping phenomena. AoI-optimal and near-optimal scheduling policies have been studied under such correlation structures, showing that carefully prioritizing less-redundant updates can significantly improve information freshness~\cite{tripathi2022optimizing,ramakanth2024monitoring}. These models typically assume fixed packet sizes and consistent source availability. In contrast, our work incorporates heterogeneous update sizes and random availability across sensing modalities, particularly capturing the intermittent access patterns of satellite-based sources, which are common in large-scale remote monitoring applications~\cite{ke2025information}.


\subsection{Summary of Contributions}
In this work, we study a remote monitoring system with heterogeneous sensing modalities including IoT sensors, UAVs, and satellites which generate updates with different sizes, availabilities, and coverage characteristics. Unlike most prior work, which typically assumes uniform update sizes, always-available sensing nodes, or non-overlapping coverage areas, we consider a setting where sensing nodes may have overlapping fields of view and intermittent availability, particularly relevant for satellite-based sensing. Our goal is to analyze fundamental trade-offs and develop constructive insights into when satellite updates should be incorporated alongside ground- and UAV-based updates.

Our contributions are summarized as follows:
\begin{itemize}[leftmargin=5mm]
    \item We derive a fundamental lower bound on the achievable average AoI in our satellite-assisted remote monitoring setting by taking into account both update sizes and availability constraints.
    \item We provide a closed-form analysis of the average AoI under stationary randomized policies, characterizing system performance both in the presence and absence of satellite-based updates. We then provide a mathematical condition that characterizes when satellite updates should be utilized, based on system parameters such as availability, transmission reliability, and packet size.
    \item We further propose an age-aware Max-Weight policy for scheduling channel access based on Lyapunov drift techniques, considering multi-packet updates with random service times.
    \item Finally, we validate our theoretical results through simulations with both idealized and trace-driven satellite availability models, demonstrating practical design implications for heterogeneous remote monitoring systems.
\end{itemize}



\section{System Model and Problem Formulation}
We consider a remote monitoring system that utilizes heterogeneous sensing modalities as illustrated in Fig.~\ref{fig:system_model}. The region of interest is discretized into a graph-based structure, where each cell can be observed and updated by one or more nodes depending on their type, location, and coverage capabilities. The central goal is to study how such a heterogeneous system can be coordinated to maintain timely and spatially distributed information using Age of Information (AoI) as the task-agnostic metric of freshness.\footnote{While different sensing platforms may capture distinct types of data (e.g., video from UAVs vs. scalar measurements from sensors), our model abstracts this heterogeneity by focusing on freshness rather than content-specific quality. This abstraction helps us provide a unified analysis of system-level information staleness and coordination.} 

\subsection{Graph-Based Network Model}
We represent the deployment region as an undirected graph $\mathcal{G} = (\mathcal{V}, \mathcal{E})$, where each vertex $\omega_m \in \mathcal{V}$ denotes a spatial cell, and each edge $(\omega_i, \omega_j) \in \mathcal{E}$ indicates adjacency for UAV mobility. The network consists of three types of nodes: 1) fixed IoT sensors $\mathcal{K}_{\text{IoT}} = \{g_1, \dots, g_N\}$, 2) mobile UAVs $\mathcal{K}_{\text{UAV}} = \{u_1, \dots, u_K\}$, and 3) satellites $\mathcal{K}_{\text{sat}}$. While we consider a constellation of satellites, they are not persistently available over the region of interest. Unlike IoT sensors and UAVs, which can provide updates continuously, satellite availability is intermittent due to orbital motion and coverage dynamics. At any given time, only a subset of satellites in the constellation may be capable of communicating with the region. For simplicity, we model the constellation abstractly as a single logical entity denoted by the subscript ``Sat.'' Let $\mathcal{K} = \mathcal{K}_{\text{IoT}} \cup \mathcal{K}_{\text{UAV}} \cup \mathcal{K}_{Sat}$ denote the set of all nodes. 

Each IoT sensor $g_i$, $i = 1,...,N$ is statically assigned to a cell through a mapping $\phi: \{g_1, \dots, g_N\} \rightarrow \mathcal{V}$, so that sensor $g_i$ covers cell $\phi(g_i)$. 
Each UAV $u_j$, $j = 1,...,K$ follows a mobility trajectory $\mathbf{p}_j(t) \in \mathcal{V}$ over time. We assume that a UAV holds its position  while transmitting an update, and only transitions to a neighboring cell once the update is successfully completed. That is, UAV $u_j$ stays at location $\mathbf{p}_j(t)$ until all packets for its current update have been delivered. Upon completion, it moves to an adjacent cell in the graph, implying that $(\mathbf{p}_j(t), \mathbf{p}_j(t+1)) \in \mathcal{E}$
if and only if an update completes at time $t$.

UAV $u_j$ can observe and update any cell within its $r_j$-hop neighborhood, which is defined as the set of all cells that are reachable from $\mathbf{p}_j(t)$ within at most $r$ hops (i.e., through a path of at most $r$ edges) in the graph $\mathcal{G} = (\mathcal{V}, \mathcal{E})$:
\begin{equation}
    \mathcal{N}^{(r_j)}(\omega_i) = \left\{ \omega_k \in \mathcal{V} \,:\, d(\omega_i, \omega_k) \le r_j \right\},
\end{equation}
where $d(\omega_i, \omega_k)$ denotes the shortest-path distance between cells $\omega_i$ and $\omega_k$ in the graph $\mathcal{G}$ measured as the number of edges along the shortest path.

Satellite availability $M(t)\in \{0,1\}$ is modeled as a two-state Markov process, which alternates between \emph{available} (A-period, $M(t)=1$) and \emph{unavailable} (U-period, $M(t)=0$) states. We assume holding times in each state are governed by geometric distributions $T_{A} \sim \text{geom}(\lambda_A)$ and $T_{U} \sim \text{geom}(\lambda_U)$. This model reflects the aggregate behavior of a satellite constellation, where the passage of multiple satellites over a given area results in stochastic coverage intervals~\cite{al2021tractable}.

At each time $t$, a node $k \in \mathcal{K}$ may be able to sense or update a cell $\omega_m \in \mathcal{V}$ depending on the node’s type, location, and coverage capability. We define a binary variable $f_{m,k}(t)$ to indicate the \emph{instantaneous update availability} of node $k$ for cell $\omega_m$ at time $t$:
\begin{equation*}
    f_{m,k}(t) =
    \begin{cases}
        1, & \text{if } \bigl(k \in \mathcal{K}_{\text{IoT}} \land \phi(k) = \omega_m\bigr) \\
          & \quad \lor \bigl(k \in \mathcal{K}_{\text{UAV}} \land \omega_m \in \mathcal{N}^{(r_k)}(\mathbf{p}_k(t))\bigr) \\
          & \quad \lor \bigl(k \in \mathcal{K}_{\text{Sat}} \land M(t)=1\bigr),\\
        0, & \text{otherwise}.
    \end{cases}
\end{equation*}


These definitions formalize the sensing capabilities of each node type, which allows us to determine which nodes are eligible to provide updates to a given cell at any time.

\subsection{Communication Model}
\label{ssec:comm}
We abstract the network’s communication layer as a shared logical wireless channel, which supports at most one transmission at any given time.\footnote{While actual implementations may involve separate frequency bands for satellites and terrestrial nodes, through our shared-channel model, we can capture a common system bottleneck such as a wireless gateway, backhaul, or processing unit where updates from heterogeneous sources ultimately contend for limited resources. This abstraction enables a unified scheduling analysis while maintaining analytical tractability.} Each node $k \in \mathcal{K}$ operates under a \emph{generate-at-will} model, i.e., a fresh update is always available whenever the node is scheduled to transmit. An update by node $k$ consists of $l_k$ packets, with each packet requiring one time slot to transmit over a single channel. Assuming no retransmissions, a complete update requires at least $l_k$ slots.

Packet-level transmission success is probabilistic: each attempt succeeds independently with probability $p_k \in (0,1]$, which reflects physical-layer conditions such as path loss, congestion, and interference. Throughout, we will assume that static IoT sensors are the most reliable due to proximity to infrastructure (e.g., fixed power or stable channels), followed by UAVs, and then satellites, which often suffer from higher path loss and handoff failures. Accordingly, we assume the ordering: $p_{g_i} > p_{u_j} > p_{\text{sat}}$ for $g_i\in \mathcal{K}_{\text{IoT}}$ and $u_j \in \mathcal{K}_{\text{UAV}}$.

A centralized controller is assumed to make all scheduling decisions based on current system information, including availability of sensing nodes, success probabilities, update sizes, and cell priorities. When the channel is idle at time $t$, the controller selects one node to begin transmission. Let $y_k(t) \in \{0,1\}$ denote the update initiation indicator, where $y_k(t) = 1$ means node $k$ initiates an update at time $t$. Let $z_k(t) \in \{0,1\}$ denote the ongoing transmission indicator. Once initiated, $z_k(t) = 1$ remains true until all $l_k$ packets are successfully transmitted or the update fails (e.g., due to satellite unavailability). The system must satisfy two constraints:
\begin{enumerate}[leftmargin=5mm]
    \item \textbf{Channel exclusivity:} Each channel can be occupied by at most one node at any given time:
    \begin{equation}
        \sum_{k \in \mathcal{K}} z_{k}(t) \le 1, \quad \forall t. \label{eq:channel_exclusivity}
    \end{equation}
    \item \textbf{Satellite transmission timing:} The satellite may initiate or continue an update only when it is available. Specifically, 
    \begin{equation}
        y_{\text{sat}}(t) = 0, \quad z_{\text{sat}}(t) = 0, \quad \text{if } M(t)=0. \label{eq:satellite_transmission}
    \end{equation}
    This ensures that no satellite transmission can occur during periods of unavailability. If the satellite becomes unavailable during an ongoing transmission, the update is immediately interrupted and considered failed. A new update initiates when the satellite becomes available again.
\end{enumerate}

\subsection{Age of Information (AoI)}

\begin{figure}
    \centering
    \includegraphics[width=0.85\linewidth]{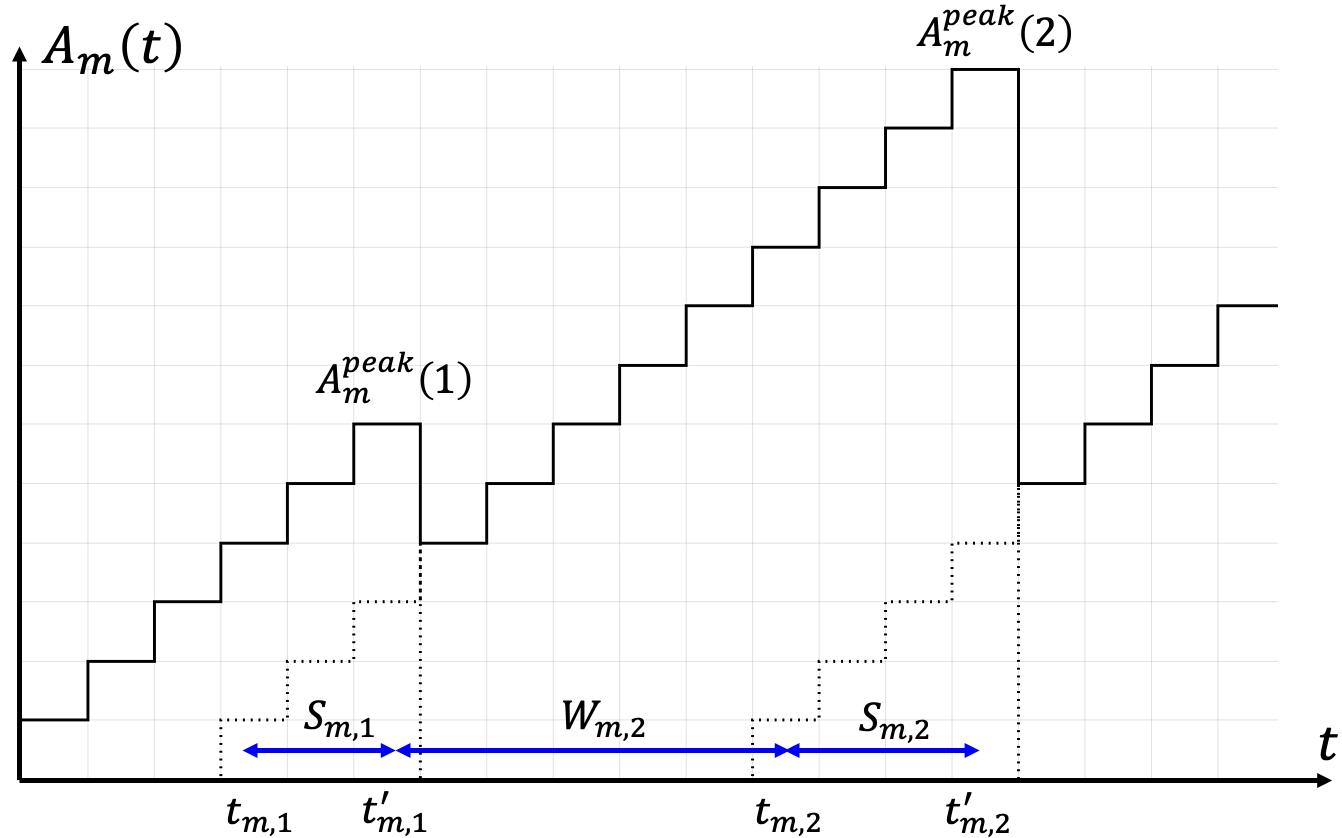}
    \caption{Age evolution at cell $\omega_m$. \vspace{-0.2in}}
    \label{fig:age_evolution}
\end{figure}

We define AoI at the monitoring destination (e.g., a fusion center) for each cell $\omega_m$. Let $A_m(t)$ denote the age of the most recently delivered information \emph{about cell $\omega_m$} available at the destination at time $t$. Thus, AoI is indexed by cells (not by sources), and it resets whenever any node successfully delivers an update that covers $\omega_m$.
We illustrate AoI evolution in Fig.~\ref{fig:age_evolution}. Recall that each update from node $k$ is considered complete only when all $l_k$ packets have been successfully transmitted and received. Let $d_k(t)\in \{0,1\}$ denote the update completion indicator for node $k$ at time $t$, where $d_k(t) = 1$ if node $k$ successfully completes an update (i.e., transmit all $l_k$ packets) at time $t$.

Let $S_k(t)=t-\tau_k^S(t)$ denote the system time of node $k$ at time $t$, where $\tau_k^S(t)$ is the generation time of the current head-of-line update in service. That is, $S_k(t)$ tracks the age of the current update in service at node $k$, which evolves as
\begin{equation}
    S_k(t+1) = \begin{cases}
        1, &\text{if } z_k(t) = 0 \text{ or } d_k(t)=1,\\
        S_k(t)+1, &\text{if } z_k(t) = 1 \text{ and } d_k(t) = 0.
    \end{cases}
\end{equation}
That is, due to the generate-at-will assumption, the system time resets to $1$ if node $k$ is idle (i.e., not transmitting) or completes the current update at time $t$; otherwise, it increases by $1$ while the update remains in service.

Then, the AoI at cell $\omega_m$ then evolves as:
\begin{equation}
    A_m(t+1) = 
    \begin{cases}
        S_k(t) + 1, &  \text{if } \exists_k : d_k(t) = 1 \\[2pt]
        & \quad \land f_{m,k}(\tau_k^S(t)) = 1,\\[4pt]
        A_m(t) + 1, & \text{otherwise}.
    \end{cases}
\end{equation}
In other words, a successful update from node $k$ contributes to resetting AoI at cell $\omega_m$ only if the node was available to cover that cell at the time the update was generated. 
The long-term average AoI at cell $\omega_m$ is given by
\begin{equation}
    \textstyle \bar{A}_m = \limsup_{T \to \infty} \frac{1}{T}  \sum_{t=1}^{T} A_m(t). 
\end{equation}

Let $t_{m,i}$ denote the generation time of the $i^{th}$ update intended for cell $\omega_m$, and $t'_{m,i}$ denote the time at which this update is successfully received. Peak AoI captures the staleness of information just before a new update is applied. For the $i^\text{th}$ update at cell $\omega_m$, the peak AoI is defined as
\begin{equation}
    A_m^{\text{peak}}(i)  := A_m(t'^{-}_{m,i}) = t'_{m,i} - t_{m,i-1}.
\end{equation}
The long-term average peak AoI at cell $\omega_m$ is then defined as
\begin{equation}
    \bar{A}_m^{\text{peak}} := \limsup_{N \to \infty} \frac{1}{N} \sum_{i=1}^N A_m^{\text{peak}}(i).
\end{equation}

\subsection{Policy Objectives}
We aim to minimize the information staleness of our heterogeneous remote monitoring system. Our focus is to determine how the diverse IoT sensor, UAV, and satellite modalities should be orchestrated to achieve timely updates across a large spatial area, despite significant differences in their reliability, coverage, and access patterns.
While AoI does not directly encode application-specific semantics, it serves as a unifying measure of freshness that is broadly applicable to sensing, tracking, and estimation tasks. Under a scheduling policy $\pi$, the Expected Weighted Sum AoI (EWSAoI) is expressed as
\begin{equation}
\mathbb{E}[J_\pi] = \limsup_{T \to \infty} \frac{1}{T |\mathcal{V}|}
\sum_{t=1}^{T} \sum_{\omega_m \in \mathcal{V}} \alpha_m
\mathbb{E}\left[A_m(t)\right],
\label{eq:obj_avg_aoi}
\end{equation}
where $\alpha_m>0$ reflects the relative importance of cell $\omega_m$.
In addition to average AoI, we also consider the Expected Weighted Sum Peak AoI (EWSPAoI), which captures the staleness of information just before each update is applied. It is defined as: 
\begin{equation} 
    \mathbb{E}[J_\pi^{\text{peak}}] = \limsup_{N \to \infty} \frac{1}{N |\mathcal{V}|} \sum_{i=1}^{N} \sum_{\omega_m \in \mathcal{V}} \alpha_m \mathbb{E}\left[A_m^{\text{peak}}(i)\right]. \label{eq:obj_peak_aoi}
\end{equation}

Directly minimizing EWSAoI/EWSPAoI in this setting is challenging due to the joint optimization over UAV mobility, satellite utilization, and scheduling decisions under stochastic network dynamics. To make the problem tractable, we first consider fixed UAV trajectories and restrict our attention to stationary randomized scheduling policies, where a node is selected for transmission according to fixed probabilities whenever the channel is idle (Sec.~\ref{sec:aoi}\&\ref{sec:rand}).
Within this policy class, we derive closed-form expressions for the EWSPAoI, enabling analytical characterization of system performance. Our analysis addresses a key question: {\it under what conditions is satellite transmission beneficial for minimizing peak AoI?} Moreover, the insights obtained from this analysis motivate our design of Lyapunov-based scheduling policies that leverage system states to further reduce AoI beyond the performance of randomized benchmarks (Sec.~\ref{sec:max-weight}).


\section{AoI Analysis: Lower Bound}
\label{sec:aoi}
We now analyze the fundamental performance limits of AoI in our heterogeneous remote monitoring system. We characterize the minimum achievable AoI under general assumptions on update sizes, transmission reliability, and availability patterns.

We begin by focusing on a single representative cell and characterize its average AoI using a renewal-theoretic approach. Each successful update received by the cell initiates a new \emph{renewal cycle}, during which the AoI increases linearly until the next update is successfully completed. This enables a closed-form characterization of both average AoI and peak AoI in terms of the inter-update intervals and service durations. We assume that updates occur in cycles indexed by $i$. We define $W_{m,i} := t_{m,i} - t'_{m,i-1}$ as the waiting time before the $i^\text{th}$ update begins transmission, representing the idle duration following the previous update. Similarly, we define $S_{m,i} := t'_{m,i} - t_{m,i}$ as the service time of the $i^\text{th}$ update, which accounts for the transmission delay from initiation to successful completion. Under these definitions, the long-term average AoI and peak AoI at cell $\omega_m$ are given as follows:
\begin{lemma} \label{lemma:aoi_expression}
    Let $W_{m,i}$ and $S_{m,i}$ be defined as above. Then, the long-term average AoI $\bar{A}_m$ and the long-term average peak AoI $\bar{A}^{peak}_m $ are expressed as
    \begin{equation}
        \textstyle\bar{A}_m = \frac{\mathbb{E}[(W_{m,i}+S_{m,i})^2]}{2\mathbb{E}[W_{m,i}+S_{m,i}]} +\frac{\mathbb{E}[S_{m,i-1}(W_{m,i}+S_{m,i})]}{\mathbb{E}[W_{m,i}+S_{m,i}]}  + \frac{1}{2},
    \end{equation}
    \begin{equation}
        \bar{A}^{peak}_m = 2\mathbb{E}[S_{m,i}]+\mathbb{E}[W_{m,i}]+1.
    \end{equation}
\end{lemma}

Within each update cycle \((t'_{m,i-1}, t'_{m,i}]\), the AoI increases linearly from \(S_{m,i-1}+1\) to \(S_{m,i-1}+W_{m,i}+S_{m,i}\), satisfying 
\(A_m(t'_{m,i-1}+s) = S_{m,i-1} + s\) for \(1 \le s \le W_{m,i} + S_{m,i}\). 
Integrating this linear growth over the cycle and applying renewal-reward theory directly yields the expressions in Lemma~\ref{lemma:aoi_expression}.\footnote{Due to space constraints, full proofs of the lemmas and propositions had to be omitted. In the appendix, we provide sketch proofs for the propositions that are given in Sec.~\ref{sec:rand}. The final version of the manuscript will link to a technical report with the full proofs of all theoretical results.}


With this result in hand, we next move to characterize the long-term behavior of node transmissions and their contribution to each cell’s information freshness. Let $\nu_{k}$ denote the long-term update frequency of node $k$ (i.e., update-level throughput), which is given by $\nu_{k} = \limsup_{T\rightarrow\infty} \frac{1}{T} \sum_{t=1}^T d_k(t)$.
Since each update from node $k$ consists of $l_k$ packets, the corresponding long-term packet-level throughput is $\nu_k l_k$. However, due to potential transmission failures, not every update initiation leads to a completed update, and hence $\sum_{t=1}^T d_k(t) \le \sum_{t=1}^T y_k(t)$.

Recall from Sec.~\ref{ssec:comm} that each update transmission is subject to per-packet success probability $p_k$. Since the $l_k$ packets of an update must all be successfully delivered, the number of required time slots to complete one update follows a negative binomial distribution. The expected number of time slots for a successful update from node $k$ is thus $l_k/p_k$. Therefore, to maintain an average update rate of $\nu_k$, node $k$ must occupy the channel for a time fraction of $\nu_k l_k/p_k$. Enforcing the channel exclusivity constraint in (\ref{eq:channel_exclusivity}), we obtain the following feasibility condition:
\begin{equation}
\sum_{k\in\mathcal{K}} \frac{l_k\nu_{k}}{p_k}\le 1.
\end{equation}

To further quantify how each node contributes to individual cells, we introduce the concept of update availability. Let $f_{m,k}$ denote the long-term fraction of time that node $k$ is available to update cell $\omega_m$. This depends on the node’s sensing characteristics: for IoT sensors, $f_{m,k}$ is binary depending on whether the sensor is assigned to $\omega_m$; for UAVs, it reflects the average fraction of time the UAV is within the sensing radius of $\omega_m$ under its trajectory; for satellites, $f_{m,k}$ captures the average visibility of $\omega_m$ given the stochastic satellite availability process $M(t)$.
Let $\mathcal{U}_{m,k}(T)$ denote the number of updates delivered by node $k$ about cell $\omega_m$ up to time $T$, and let $\mathcal{D}_{m,k}(T)$ denote the corresponding number of successfully delivered packets. These metrics respectively reflect update-level and packet-level contributions to each cell. Then, the long-term update-level throughput from node $k$ regarding cell $\omega_m$ is defined as  
\begin{equation}
\gamma_{m,k} = \limsup_{T\rightarrow\infty} \frac{\mathcal{U}_{m,k}(T)}{T}= \nu_{k} f_{m,k}. \label{eq:act_freq}
\end{equation}

With these quantities in place, we now present a performance bound on the achievable AoI. The following result provides a universal lower bound on the expected weighted sum AoI under any non-anticipatory scheduling policy $\pi$.
\begin{lemma} \label{lemma:lower_bound}
    For a heterogeneous remote monitoring system characterized by parameters $\{\mathcal{G},\alpha_m,p_k,l_k\}$, the following lower bound holds for any non-anticipatory policy $\pi$:
    \begin{align}
        &\mathbb{E}[J_\pi] \ge \frac{1}{2|V|}\sum_{\omega_m \in V} \alpha_m\\
        &+ \frac{1}{2|V|^2} \left(\sum_{\omega_m \in V} \sqrt{\frac{\alpha_m \sum_{k\in \mathcal{K}_m}l_k\gamma_{k,m}}{ (\sum_{k\in \mathcal{K}_m}p_kf_{k,m})(\sum_{k\in \mathcal{K}_m} \gamma_{k,m})}}\right)^2. \nonumber
    \end{align}
\end{lemma}
\noindent where $l_{\text{min},m} = \min_{k\in\mathcal{K}_m}l_k$.

This result reveals the structural dependencies of AoI performance on update sizes ($l_k$), success probabilities ($p_k$), and spatial-temporal availability ($f_{m,k}$). The lower bound can be obtained from Cauchy-Schwarz inequality, Jensen's inequality and the communications constraint (\ref{eq:act_freq}).

\section{Randomized Scheduling Policy}
\label{sec:rand}
In this section, we analyze the performance of a heterogeneous remote monitoring system under a stationary randomized scheduling policy. Specifically, we derive closed-form expressions for the peak AoI under such policies, first in the absence of satellite support (Sec.~\ref{ssec:nosat}) and then in the presence of a probabilistically available satellite link (Sec.~\ref{ssec:sat}). This framework provides a baseline for understanding the impact of node heterogeneity, spatial dynamics, and intermittent connectivity on information freshness.

\subsection{Randomized Scheduling without Satellite Support}
\label{ssec:nosat}

In the absence of a satellite link, the scheduling policy involves only IoT sensors and UAVs. Let $\mathcal{K}_{*} = \mathcal{K}_{IoT} \cup \mathcal{K}_{UAV}$ be the set of IoT sensors and UAVs. The scheduler employs a stationary randomized scheme where each node $k \in \mathcal{K}_{*}$ is granted access to the channel with a fixed probability $\mu_k$, subject to the constraint $\sum_{k \in \mathcal{K}_{*}} \mu_k \leq 1$.
When a node $k$ is scheduled, it transmits an update consisting of $l_k$ packets, each occupying one time slot. Each packet is transmitted successfully with probability $p_k$ (independently across packets). In particular, if the first packet transmission is successful (which occurs with probability $\mu_k p_k$), the node holds the channel until all $l_k$ packets are transmitted successfully. Otherwise, if the first transmission fails (with probability $\mu_k (1-p_k)$), the channel becomes idle immediately. Under the generate-at-will assumption, every node always has a fresh update ready whenever it is scheduled.

The following proposition provides expressions for the weighted average AoI and the weighted average peak age under the randomized scheduling policy without satellite support. 

\begin{proposition} \label{proposition:without_sat}
For the remote monitoring system without satellite support, under the randomized scheduling policy, the EWSPAoI is given by
\begin{equation*}
\mathbb{E}[J^{\text{peak}}_{\pi}] = \frac{1}{|V|} \sum_{\omega_m \in V} \frac{\alpha_m}{B_m} \Bigl(1+\sum_{k\in\mathcal{K}_*} \mu_k(f_{m,k}+1) (l_k-1)\Bigr) + 1,
\end{equation*}
where $B_{m}= \sum_{k \in \mathcal{K}_*} \mu_k p_k f_{m,k}$.
\end{proposition}

The above closed-form expressions quantify the impact of scheduling probabilities, transmission reliability, and coverage availability on information freshness at each cell. The term $B_m$ captures the effective update rate for cell $\omega_m$, i.e., how frequently the cell receives a successful update from a node that covers it. Larger $B_m$ implies more frequent refreshing of the cell’s information, which leads to lower AoI. The expression for peak age $\mathbb{E}[J^{\text{peak}}_{\pi}]$ highlights the role of both access probability and update size. A sketch proof of this proposition is provided in Appendix~\ref{appendix:without_sat}.

\subsection{Randomized Scheduling with Satellite Support}
\label{ssec:sat}
We now extend the analysis to include satellite support, where a satellite node can intermittently participate in the update process due to its time-varying visibility. In this scenario, the scheduler allocates channel access to both terrestrial nodes and the satellite using fixed probabilities. When the satellite is available, it can cover all cells simultaneously through one update. However, the success of a satellite update depends not only on channel reliability but also on whether the satellite remains available long enough to complete the transmission of all required packets.
The following proposition characterizes the system performance in this setting by providing closed-form expressions for the EWSPAoI in terms of system parameters including the satellite's availability and transmission probability. 
\begin{proposition} \label{proposition:with_sat}
For the satellite-assisted remote monitoring system, under the randomized scheduling policy (with satellite support), the EWSPAoI is  given by
\begin{align*}
    \mathbb{E}[J^{peak}_{\pi}] &\textstyle= \frac{1}{|V|} \sum_{\omega_m\in V} \alpha_m (2\mathbb{E}[S_m]+\mathbb{E}[W_m]+1),
\end{align*}
where
\begin{align*}
    &\textstyle \mathbb{E}[S_m] 
    \textstyle= \frac{1}{C_m}\hspace{-0.05cm} \left(\hspace{-0.05cm}\sum_{k\in\mathcal{K}_*} \hspace{-0.1cm}\mu_k f_{m,k}(l_k-1) \hspace{-0.05cm}+\hspace{-0.05cm} \frac{\beta_{sat}\mu_{sat} p_{sat} \gamma (l_{sat}-1) }{p_{sat} + \lambda_A - p_{sat}\lambda_A}\hspace{-0.05cm}\right)\hspace{-0.05cm}, \nonumber \\
    &\mathbb{E}[W_m] \textstyle = \frac{1}{C_{m}} \big(1+ \sum_{k\in\mathcal{K}_*} \mu_k (1-f_{m,k})(l_k-1) \\
        &\hspace{4cm}\textstyle + \beta_{sat}\mu_{sat}p_{sat}(1-\gamma)\mathbb{E}[X_{w}]\big),
\end{align*}
with $C_m = \sum_{k\in\mathcal{K}_*}\mu_k p_kf_{m,k} + \beta_{sat}\mu_{sat}p_{sat}\gamma$,
\begin{gather*}
    \textstyle \gamma = \left(\frac{p_{sat}(1-\lambda_A)}{p_{sat}+\lambda_A -p_{sat}\lambda_A}\right)^{l_{sat}-1}, 
\end{gather*}
\begin{equation*}
     \beta_{\text{sat}} = 
    \frac{D_A(\bar\lambda_A)(1-D_U(\bar\lambda_U))}
     {D_A(\bar\lambda_A)(1-D_U(\bar\lambda_U)) + D_U(\bar\lambda_U)(1-D_A(\bar\lambda_A))}, 
\end{equation*}
\begin{align*}
D_A(q) = \sum_{k \in \mathcal{K}} \mu_k \, \psi(q; \ell_k, p_k), ~ D_U(q) =  \sum_{k \in \mathcal{K}_{\text{*}}} \mu_k \, \psi(q; \ell_k, p_k),
\end{align*}
\begin{equation*}
\textstyle\psi(q; \ell, p) = q \left( 1 - p + p \left( \frac{p q}{1 - (1-p)q} \right)^{\ell - 1} \right),
\end{equation*}
%
\begin{equation}
        \textstyle\mathbb{E}[X_{w}] = \frac{\sum_{k=0}^\infty k\lambda_A(1-\lambda_A)^{k}(1-I_{p_{sat}}(l_{sat}-1,k-l_{sat}+2))}{\sum_{k=1}^\infty \lambda_A(1-\lambda_A)^{k}(1-I_{p_{sat}}(l_{sat}-1,k-l_{sat}+2))} + 1, \nonumber
\end{equation}
where $\bar\lambda_A=1-\lambda_A$, $\bar\lambda_U = 1-\lambda_U$ and  $I_p(a,b)$ is the regularized incomplete Beta function.
\end{proposition}
These expressions capture how the satellite's intermittent availability and transmission reliability impact the average system performance. Notably, the parameter $\gamma$ plays a key role in quantifying the success probability of satellite-assisted updates within the available time window, while $X_w$ captures the expected number of time slots wasted due to satellite update failures. A sketch proof can be found in Appendix~\ref{appendix:with_sat}.

\subsection{IoT-Satellite Scenario} \label{subsec:iot_sat}
In order to gain insight and provide tractable analysis into when satellite support is beneficial, we first focus on a simplified IoT-Satellite setting where UAVs are excluded. Each cell is covered by an IoT sensor, and without loss of generality, the index of a cell is identical to the index of its associated sensor, i.e., $k = \phi(k)$. We assume that all cells are equally important, i.e., $\alpha_m = 1$ for all $\omega_m$.

Further, directly comparing the age performance between the scenarios with and without satellite support is intractable due to the complexity of the expressions involved, as seen in Propositions~\ref{proposition:without_sat} and \ref{proposition:with_sat}. In particular, the expected number of time slots wasted due to satellite update failures, denoted by $\mathbb{E}[X_w]$, involves the regularized incomplete Beta function $I_p(a,b)$, which complicates analytical comparison. To simplify the analysis, we assume that the durations of A-period and U-period are very large, i.e., $\lambda_A$ and $\lambda_U$ are very small. Under this assumption, the effect of satellite update failures becomes negligible and can be ignored. Then, we have $\gamma \rightarrow1$, $\mathbb{E}[X_w]\rightarrow 0$ and $\beta_{\text{sat}}\rightarrow \frac{\lambda_U}{\lambda_A + \lambda_U}$ as $(\lambda_A,\lambda_U)\rightarrow (0,0)$.

During the U-period, each sensor $k$ accesses the channel with probability $\mu_k$, where the probabilities satisfy $\sum_{k\in\mathcal{K}_{IoT}} \mu_k = 1$. During the A-period, the satellite accesses the channel with probability $\mu_{\mathrm{sat}} = \alpha$, while each sensor $k$ accesses the channel with probability $(1 - \alpha)\mu_k$. Using these access probabilities, the expected service time and waiting time for cell $m$ are given by
\begin{align*}
     \mathbb{E}[S_m] &\textstyle= \frac{\lambda_A\mu_m(l_m-1) + \lambda_U(\alpha(l_{\text{sat}}-1) + (1-\alpha)\mu_m(l_m-1))}{\lambda_A \mu_m p_m + \lambda_U (\alpha p_{\text{sat}} + (1-\alpha)\mu_m p_m)}, \\
     \mathbb{E}[W_m] &\textstyle= \frac{\lambda_A(\mu_m + \sum_{k\neq m}\mu_kl_k) + \lambda_U(\alpha+(1-\alpha)(\mu_m + \sum_{k\neq m} \mu_k l_k))}{\lambda_A \mu_m p_m + \lambda_U (\alpha p_{\text{sat}} + (1-\alpha)\mu_m p_m)}.
\end{align*}
By substituting these expressions into $\mathbb{E}[J^{peak}_{\pi}]$ rearranging terms, the problem can be formulated as:
\begin{gather*}
\min_\alpha J(\alpha) = \sum_{m=1}^M \frac{N_{0,m}+N_{1,m}\alpha}{D_{0,m} + D_{1,m}\alpha},\\
    \text{where } N_{0,m} = 2\lambda_A[(l_m-1)+M_m], \quad \textstyle M_m = \sum_{k\neq m} \mu_k, \\
    N_{1,m}=2\lambda_U[(l_{\text{sat}}-1)-\mu_m(l_m-1)-M_m], \\
    D_{0,m} = \lambda_A \mu_m p_m, \quad D_{1,m} = \lambda_U [p_{\text{sat}}-\mu_m p_m].
\end{gather*}
Although $J(\alpha)$ is not itself the expected peak age, minimizing it is equivalent to minimizing $\mathbb{E}[J^{peak}_{\pi}]$. 
We have $J'(\alpha) = \textstyle \sum_{m=1}^M \frac{K_m}{(D_{0,m} + \alpha D_{1,m})^2}$,
where $K_m = N_{1,m}D_{0,m} - N_{0,m}D_{1,m}$ measures the relative advantage of satellite updates over IoT sensor $m$.  If $K_m > 0$, the satellite is more effective, and increasing $\alpha$ reduces the age; if $K_m < 0$, the sensor is preferable, and increasing $\alpha$ may increase the age.  Since $J'(\alpha)$ sums these terms, its sign determines whether $J(\alpha)$ decreases or increases with $\alpha$. When all $K_m$ share the same sign, the optimum lies at $\alpha^\star = 1$ (all positive) or $\alpha^\star = 0$ (all negative). If the $K_m$ have mixed signs, an interior optimum $0 < \alpha^\star < 1$ may exist, balancing satellite and sensor benefits. For instance, when $M=2$, $\alpha^\star = \frac{\sqrt{K_1}D_{0,2} - \sqrt{K_2}D_{0,1}}{\sqrt{K_2}D_{1,1} - \sqrt{K_1}D_{1,2}}$.

\section{Max-Weight Policy}
\label{sec:max-weight}
In this section, we design a scheduling strategy for the IoT–UAV–satellite remote monitoring system that minimizes the AoI while ensuring long-term stability. Unlike existing Lyapunov-based policies~\cite{kadota2019minimizing,kadota2019scheduling} which assume single-packet transmissions and make decisions at every slot, our setting involves multi-packet updates with random service times that keep the channel occupied for variable durations. Let $\tau_i$ denote the $i^{\text{th}}$ epoch when the channel becomes idle and a new scheduling decision is made. Since transmissions may span multiple slots, scheduling decisions are taken only at these epochs $\{\tau_i\}$. For each node $k$, let $\mathcal{M}_{\tau_i}(k)$ denote the set of cells it refreshes if selected at $\tau_i$, and let $A_m(\tau_i)$ be the AoI of cell $\omega_m$ at that epoch.

To track the service level of each node, we introduce the throughput debt of node $k$ at epoch $\tau_i$ as $x_k(\tau_i) = i\,\bar \nu_k - \sum_{s=1}^{\tau_i} d_k(s)$,
where $\bar \nu_k$ is the long-term update-level target throughput for node $k$. The term $i \bar \nu_k$ represents the minimum number of updates that node $k$ should have delivered by epoch $\tau_i$, while $\sum_{s=1}^{\tau_i} d_k(s)$ is the actual number delivered. The positive part, $x_k^+(\tau_i) = \max\{x_k(\tau_i),0\}$, serves as a virtual queue that guides scheduling decisions and ensures stability. Strong stability of this virtual queue, i.e., $\lim_{T\to\infty} \frac{1}{T} \sum_{i=1}^{T} \mathbb{E}[x_k^+(\tau_i)] < \infty$, is sufficient to guarantee that the long-term throughput of node $k$ satisfies $\nu_k \geq \bar \nu_k$, which ensures that its update rate meets or exceeds the target~\cite{neely2010stability}. Given the state $\mathbb{S}(\tau_i) := \bigl( \{A_m(\tau_i)\}_{m \in \mathcal{V}}, \{x_k^+(\tau_i)\}_{k \in \mathcal{K}} \bigr)$, we define the quadratic Lyapunov function as
\begin{equation}
    L(\mathbb{S}(\tau_i)) = \sum_{\omega_m \in \mathcal{V}} \alpha_m A_m^2(\tau_i)
          + \frac{\beta}{2} \sum_{k \in \mathcal{K}} \bigl(x_k^+(\tau_i)\bigr)^2,
\end{equation}
where $\beta>0$ controls the importance on throughput debt. The Lyapunov drift is then
\begin{equation}
    \Delta(\mathbb{S}(\tau_i)) := \mathbb{E} \bigl[ L(\tau_{i+1}) - L(\tau_i) \,\big|\, \mathbb{S}(\tau_i) \bigr].
\end{equation}
Using the update dynamics of $A_m(\tau_i)$ and $x_k^+(\tau_i)$, the one-step Lyapunov drift can be upper-bounded as
\begin{equation}
    \Delta(\mathbb{S}(\tau_i)) \le B(\tau_i) 
    - \sum_{k \in \mathcal{K}}\mathbb{E}[u_k(\tau_i)| \mathbb{S}(\tau_i)] C_k(\tau_i), 
    \label{eq:drift_upperbound}
\end{equation}
where $B(\tau_i) = \beta |\mathcal{K}|/2 + \beta \sum_{k \in \mathcal{K}} \bar \nu_k x_k^+(\tau)$ and 
\begin{align*}
        C_k(\tau_i) &= \beta p_k x_k^+(\tau_i) \\
        &+ p_k \hspace{-0.4cm}\sum_{\omega_m \in \mathcal{M}_{\tau_i}(k)} \hspace{-0.4cm}\textstyle \alpha_m A_m(\tau_i)\left(A_m(\tau_i) + 2-\frac{2}{p_k}\right) \nonumber \\
        &- 2 l_k\hspace{-0.4cm} \sum_{\omega_m \notin \mathcal{M}_{\tau_i}(k)} \hspace{-0.4cm} \alpha_m  A_m(\tau_i) - {\textstyle \frac{l_k(l_k+p_k-1)}{p_k}} \hspace{-0.15cm} \sum_{\omega_m \in \mathcal{V}}\alpha_m.
\end{align*}

In contrast to one-slot transmission models where the uncovered cells' ages increase deterministically by one, our setting involves multi-packet updates with random service times. The duration of the busy period when node $k$ is scheduled  is a random variable. Therefore, unlike in single-slot Lyapunov formulations, the random epoch length introduces decision-dependent penalties for uncovered cells and modifies the benefit from covered cells, both of which appear explicitly in $W_k(\tau_i)$. This ensures that the scheduling policy accounts not only for age reduction but also for the expected age growth of other cells during the potentially long busy period.

The choice of the target throughput $\bar{\nu}_k$ for each node $k$ is critical since it directly determines the long-term service guarantees enforced by the virtual queues. To determine appropriate values for $\bar{\nu}_k$, we exploit the fact that the long-term average AoI of cell $\omega_m$ admits a lower bound that is inversely proportional to its total update rate $\sum_{k \in K_m} \nu_k f_{m,k}$. This bound reflects the principle that more frequent updates provably reduce the minimum achievable AoI. Using this bound as a surrogate objective, the desired throughput targets $\bar{\nu}_k$ can be obtained by solving
\begin{equation}
    \min_{\{\nu_k\}} \quad 
    \sum_{\omega_m \in \mathcal{V}} 
    \frac{\alpha_m}{\sum_{k \in \mathcal{K}_m} \nu_k f_{m,k}},
    \label{eq:nu_optimization}
\end{equation}
subject to $\sum_{k\in \mathcal{K}} \frac{l_k \nu_k}{p_k} \le 1$. The resulting optimal values $\bar{\nu}_k$ serve as the reference rates in the virtual queues.

In the special case with no UAV or satellite support, where each cell $\omega_m$ is covered by a unique node $k=m$ (i.e., $\mathcal{K}_m = \{m\}$), \eqref{eq:nu_optimization} simplifies to minimizing $\sum_{m\in\mathcal{V}} \alpha_m /(2\nu_m)$ subject to $\sum_{k} l_k \nu_k / p_k \le 1$. Solving this convex problem via the KKT conditions yields a closed-form solution $\bar \nu_k = \frac{\sqrt{\alpha_k p_k / l_k}}{\sum_j \sqrt{\alpha_j l_j / p_j}}$.

\section{Numerical Results}

In this section, we present simulation results to verify our analytical insights and policy designs, and to explore the conditions under which satellite-assisted updates improve network performance. We evaluate the average peak AoI in a heterogeneous network where IoT sensors and a UAV share access to the channel with optional satellite support. 

\subsection{Simulation Environment}
We simulate a heterogeneous network consisting of terrestrial IoT sensors, a UAV, and satellites. The network is arranged on a $4\times4$ grid with $16$ cells, where $6$ IoT sensors are uniformly deployed and a single UAV patrols with a coverage radius of one hop ($r=1$). Under the terrestrial-only setting, channel access is equally shared among the $7$ terrestrial nodes, giving each node an access probability $\mu_k = 1/7$. When satellite support is enabled, the satellite is prioritized with an access probability $\mu_{sat} = 0.5$, and the remaining terrestrial nodes share the residual probability $\mu_k = 1/14$. The IoT sensors generate updates with packet length $l_{\text{IoT}} = 2$ and success probability $p_{\text{IoT}} = 0.8$, while the UAV uses $l_{\text{uav}} = 6$ and $p_{\text{uav}} = 0.8$.

\begin{figure}
    \centering
    \begin{subfigure}[t]{0.49\linewidth}
        \centering
        \includegraphics[width=\linewidth]{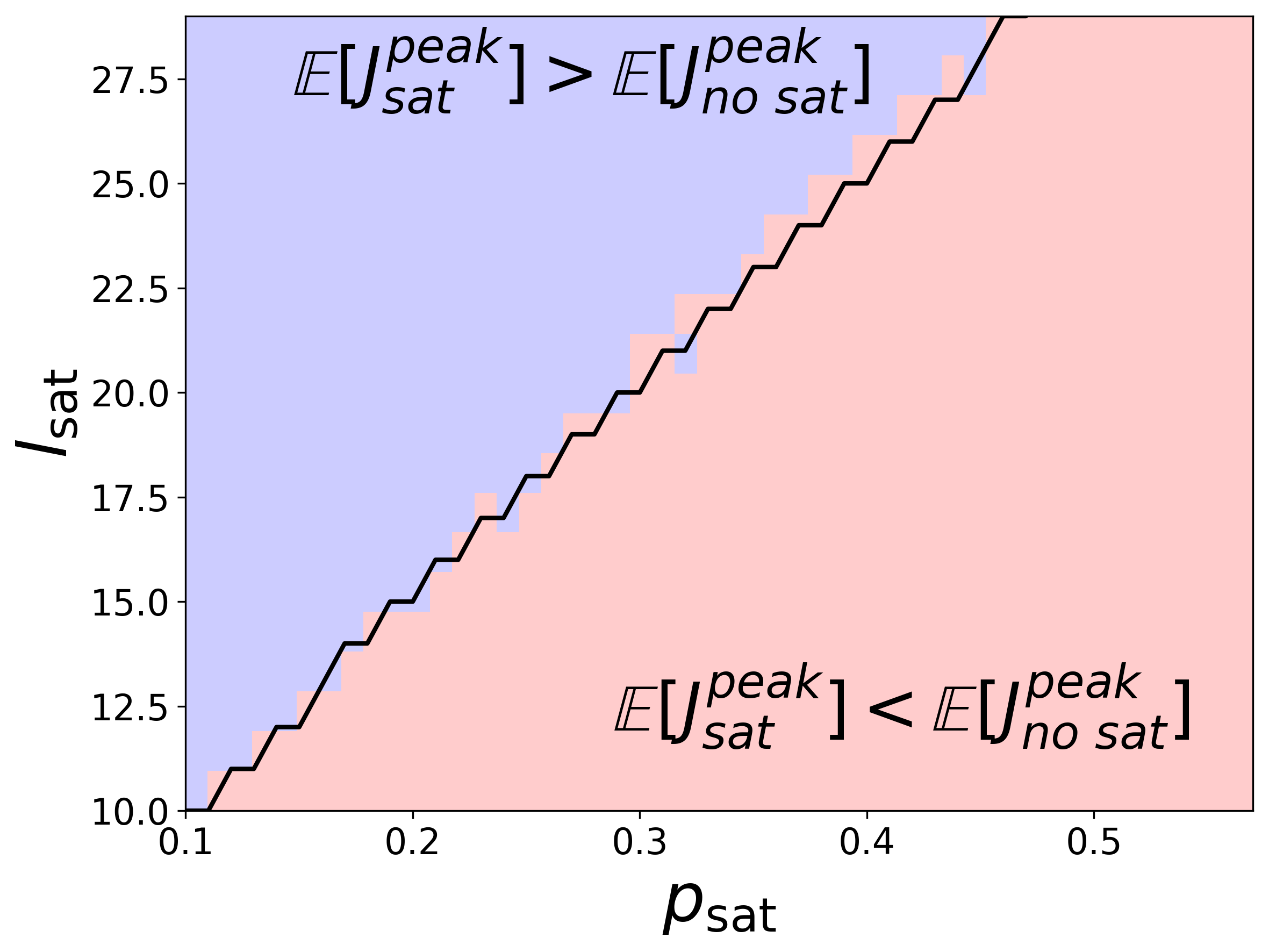}
        \caption{Idealized geometric availability model.}
    \end{subfigure}
    \hfill
    \begin{subfigure}[t]{0.49\linewidth}
        \centering
        \includegraphics[width=\linewidth]{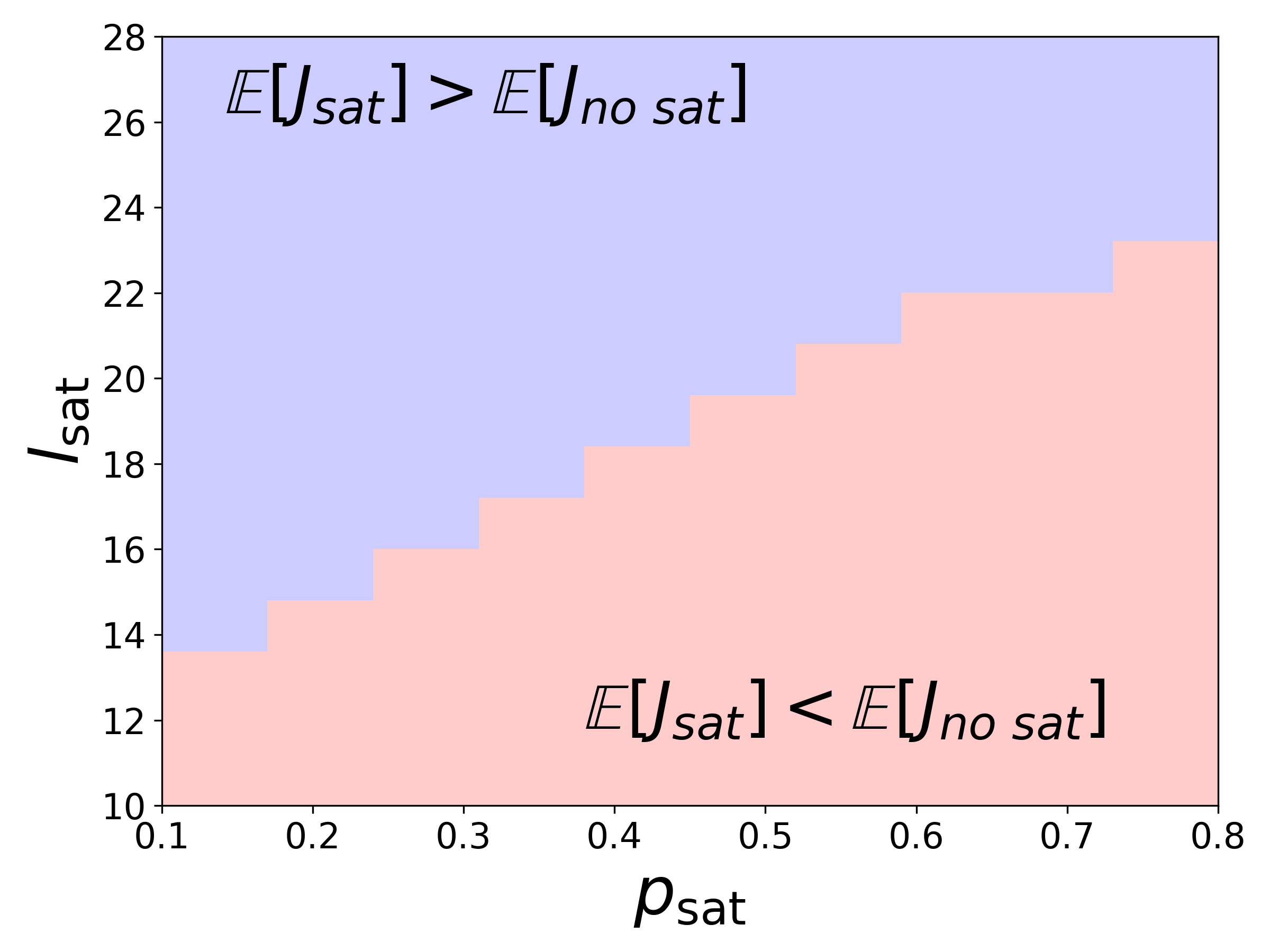}
        \caption{Trace-driven Walker-Star availability model.}
    \end{subfigure}
    \caption{Parameter space for satellite-assisted update performance with $p_{sat}$ (horizontal) and $l_{sat}$ (vertical). Red regions indicate lower average AoI with satellite support.}
    \label{fig:p_l_compare}
\end{figure}

We vary the satellite packet length $l_{sat}$, success probability $p_{sat}$, and availability parameter $\lambda_A$ to examine how these factors jointly influence the average peak AoI. In the default case, we set $l_{sat} = 20$ and $p_{sat} = 0.6$. Two models are used to characterize satellite availability:
\begin{itemize}[leftmargin=5mm]
    \item In the \emph{idealized geometric model}, the satellite alternates between available and unavailable states according to transition rates $\lambda_A$ and $\lambda_U = 0.05$.
    
    \item In addition, a \emph{trace-driven model} is considered, which uses real-world visibility data from the $12$-satellite Walker-Star constellation ($3$ orbital planes, $550$ km altitude) sampled at $1$-second resolution~\cite{walkerstar,han2024cooperative}. This trace reveals an average availability of $59.7\%$, with mean on/off durations of $967.1$ seconds and $657.0$ seconds, respectively.
\end{itemize}
All numerical results are averaged over $20$ runs.

\subsection{Results and Discussion}

We first evaluate how the satellite's transmission success probability $p_{sat}$ and update length $l_{sat}$ jointly affect the network’s average peak AoI. Figures~\ref{fig:p_l_compare}(a) and (b) show the parameter space for satellite-assisted updates under the idealized and Walker-Star trace-driven models, respectively. In both figures, red regions indicate parameter combinations where satellite assistance improves AoI compared to terrestrial-only operation, while blue regions correspond to cases where satellite support worsens AoI. For the idealized model, the satellite-assisted scheme is most beneficial when $p_{sat}$ is high and $l_{sat}$ is small. The black line represents the theoretical boundary separating these regions, obtained numerically by solving the equations in Propositions~\ref{proposition:without_sat}\&\ref{proposition:with_sat} (as a closed-form expression is not available), and closely matches the simulation. The trace-driven results confirm the trend in (a), with satellite support remaining advantageous only when transmission reliability is sufficiently high and update sizes are moderate.

\begin{figure}
    \centering
    \begin{subfigure}[t]{0.48\linewidth}
        \centering
        \includegraphics[width=\linewidth]{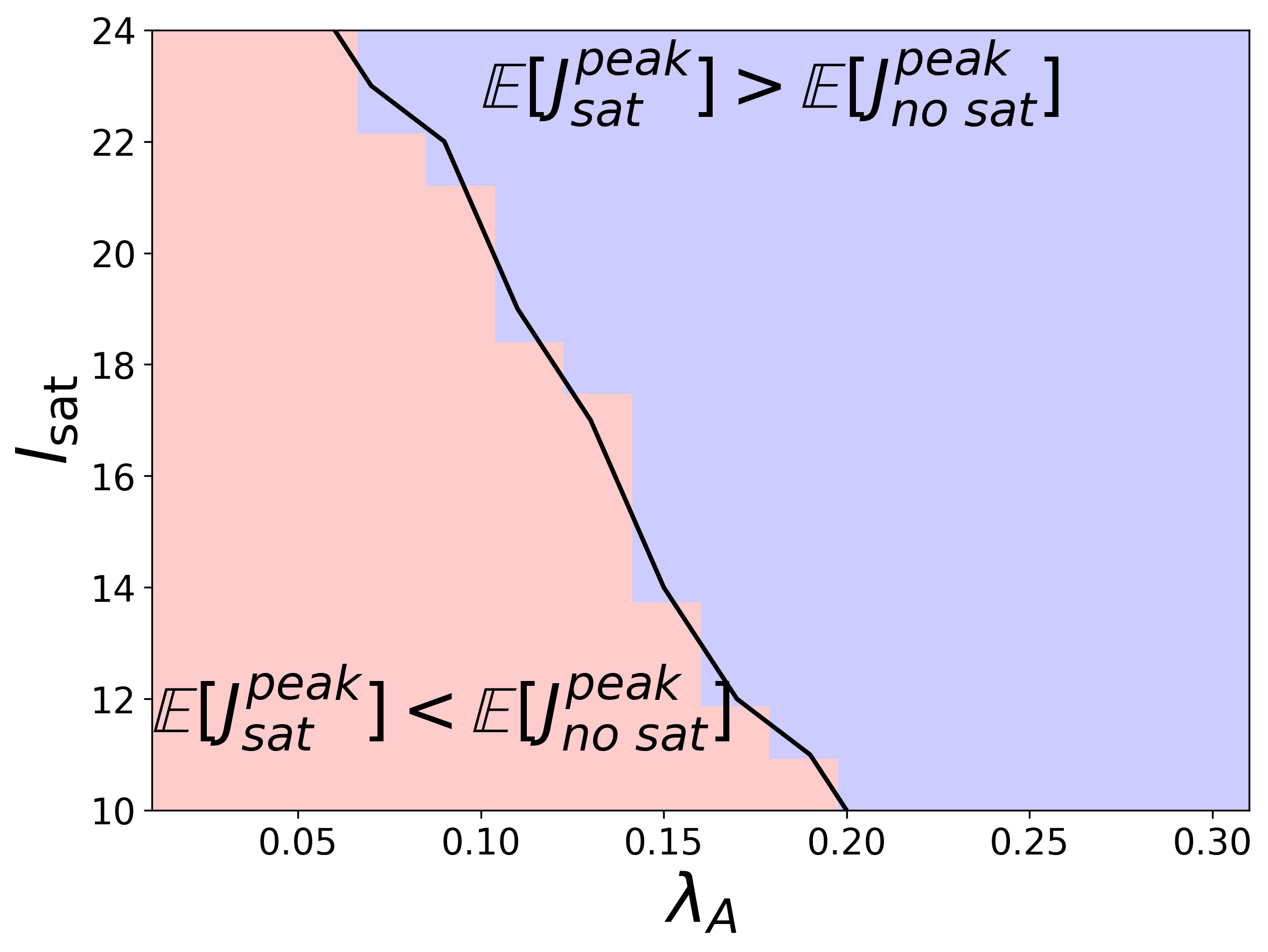}
        \caption{Impact of availability parameter $\lambda_A$ and $l_{sat}$.}
    \end{subfigure}
    \hfill
    \begin{subfigure}[t]{0.48\linewidth}
        \centering
        \includegraphics[width=\linewidth]{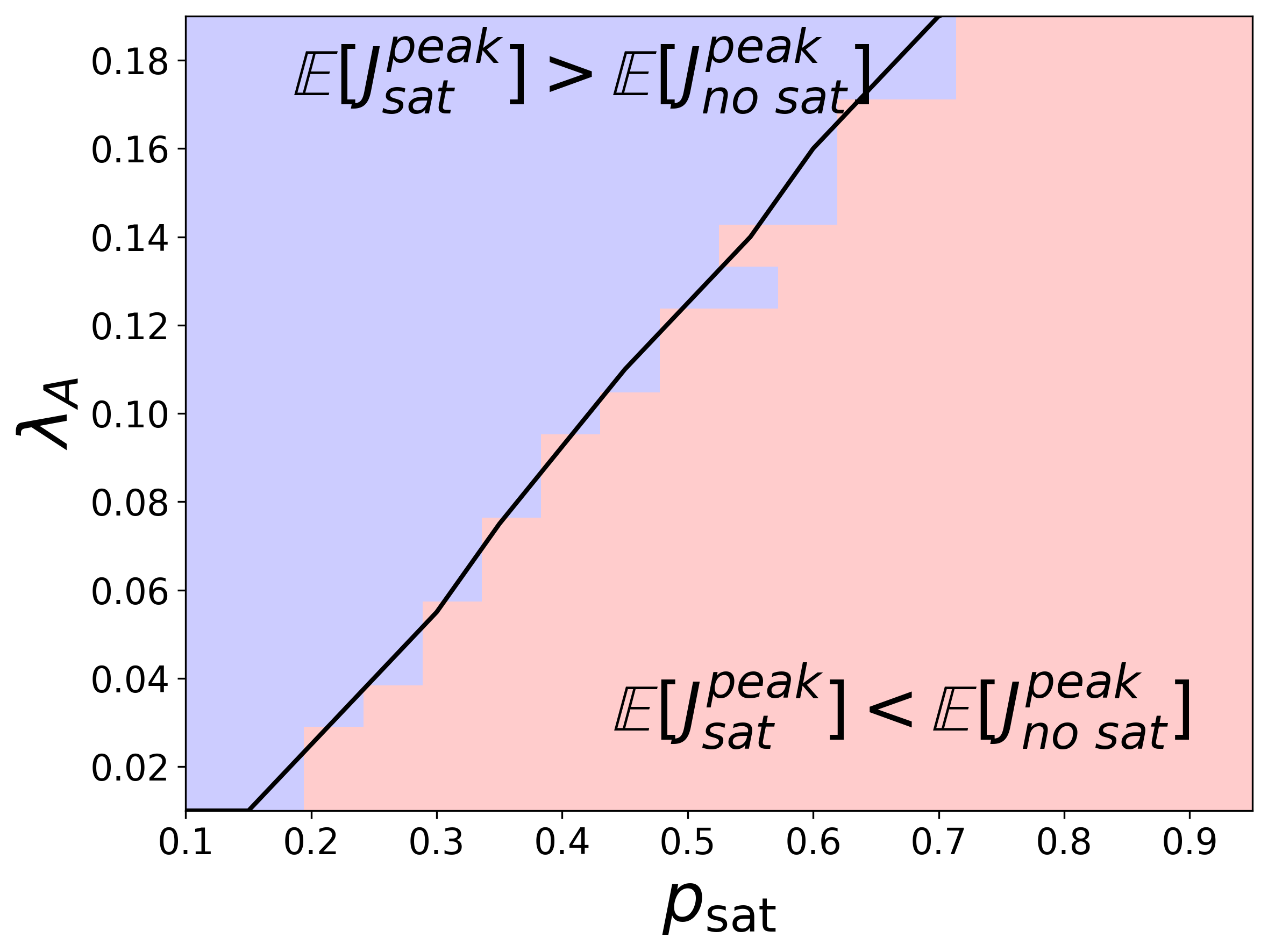}
        \caption{Impact of availability parameter $\lambda_A$ and $p_{sat}$.}
    \end{subfigure}
    \caption{Parameter space for satellite-assisted update performance under the idealized model. As in Fig.~\ref{fig:p_l_compare}, red regions indicate lower average AoI with satellite support. \vspace{-0.1in}}
    \label{fig:lambda_effect}
\end{figure}

Next, we investigate how satellite availability interacts with packet length and transmission success probability under the idealized model. 
Figure~\ref{fig:lambda_effect}(a) shows the impact of $\lambda_A$ and $l_{sat}$, which demonstrates that longer satellite availability periods (smaller $\lambda_A$) reduce AoI when update sizes are moderate. This is because a smaller $\lambda_A$ lowers the fraction of updates that fail due to the satellite becoming unavailable during transmission, thereby enhancing the effectiveness of satellite-assisted updates.  Figure~\ref{fig:lambda_effect}(b) examines the combined effect of $\lambda_A$ and $p_{sat}$, which confirms that the satellite’s benefit is maximized when both reliability and availability are high.

\begin{figure}
    \centering
    \includegraphics[width=0.9\linewidth]{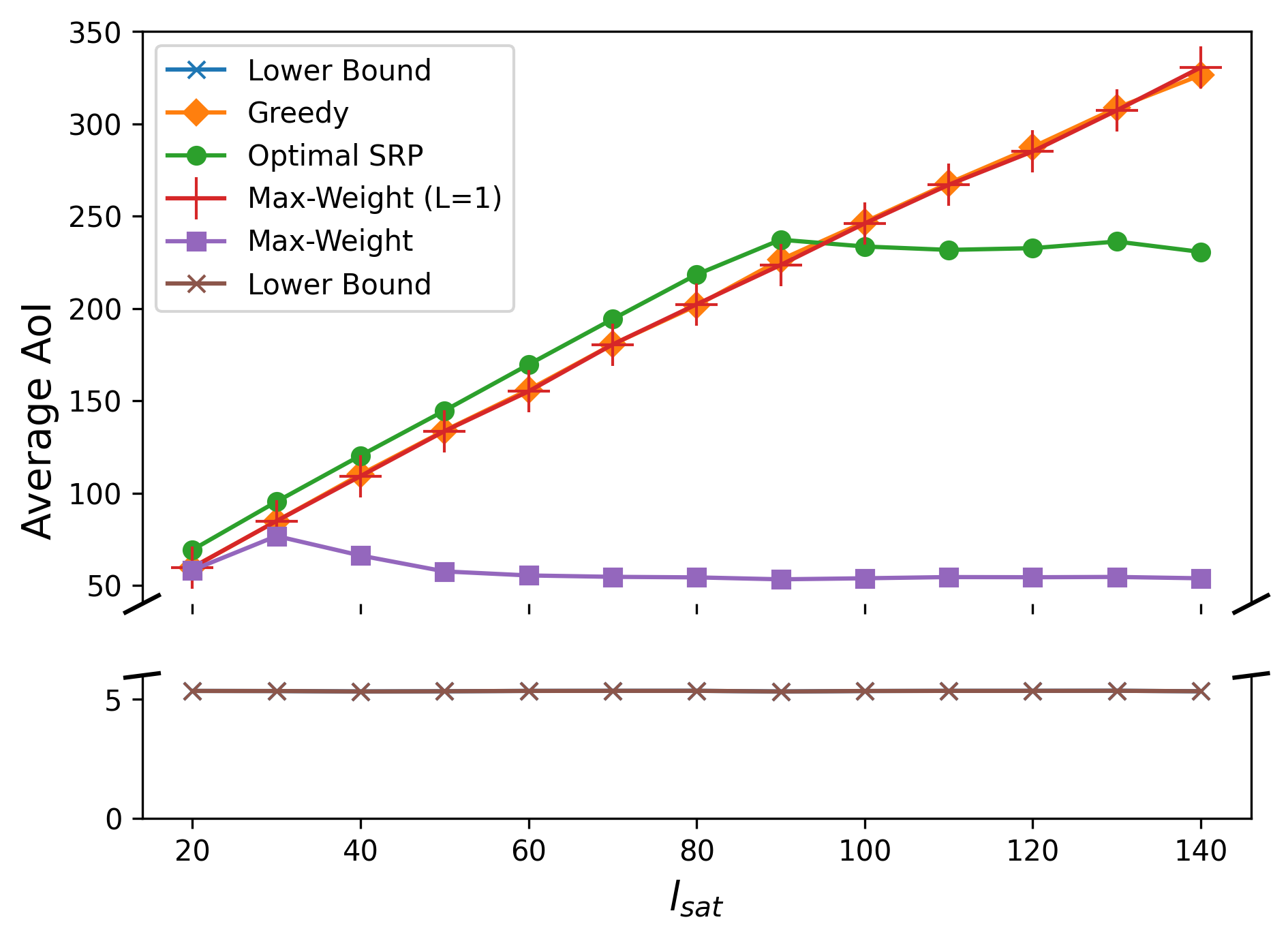}
    \caption{Average AoI achieved by the optimal stationary randomized (SR) and max-weight (MW) policies, as well as baseline policies and our theoretical lower bound. \vspace{-0.2in}}
    \label{fig:age_SRP_MW2}
\end{figure}

Figure~\ref{fig:age_SRP_MW2} compares the average AoI achieved by different scheduling policies as the satellite packet length $l_{\text{sat}}$ increases. The Greedy policy selects the transmitter that maximizes the instantaneous age reduction, given by $\arg\max_k \sum_{m \in \mathcal{M}_k} \alpha_m A_m(t)$. The Max-Weight policy with $L=1$ (MWL1)~\cite{kadota2018scheduling} follows the rule $\arg\max_k \sqrt{p_k} \sum_{m \in \mathcal{M}_k} \sqrt{\alpha_m} A_m(t)$, which balances the current age with the transmission reliability. 
We see that for small $l_{\text{sat}}$, the satellite’s full coverage dominates, and all policies achieve low AoI. As $l_{\text{sat}}$ grows, however, the satellite's long transmission time outweighs this benefit, causing Greedy and MWL1 to degrade, while our MW policy limits satellite usage.

\section{Conclusion}
In this work, we presented an analytical framework for heterogeneous remote monitoring systems that integrate IoT sensors, UAVs and satellite-based updates. Our analysis revealed fundamental trade-offs between update sizes, node availability and transmission reliability. We derived closed-form expressions and mathematical conditions that identify when satellite support improves the age performance, and employed this to develop optimal stationary random and max-weight scheduling policies. Simulation results validated our theoretical findings and highlighted the significance of satellite availability, update length and transmission reliability in determining overall system performance. Future work will explore near-optimal wireless scheduling and UAV mobility planning strategies that further utilize these insights to improve information freshness in dynamic environments.

\balance
\bibliographystyle{IEEEtran}
\bibliography{Sunjung_satellite}

\appendix

\subsection{Proof of Lemma~\ref{lemma:aoi_expression}}\label{appendix:aoi_expression}

    Let us define the $i^{th}$ update cycle at cell $\omega_m$ as the interval between two consecutive successful updates, i.e., the interval $(t'_{m,i-1},t'_{m,i}]$. Within each update cycle, the AoI grows linearly from $S_{m,i-1}+1$ to $S_{m,i-1}+W_{m,i}+S_{m,i}$. More explicitly, at any time $t=t'_{m,i-1}+s$ within the $i^{th}$ cycle,
    \begin{equation}
        A_m(t'_{m,i-1}+s) = S_{m,i-1} +s, ~\text{ for } 1\le s \le W_{m,i}+S_{m,i}.
    \end{equation}

    Summing over the inter-update period, we have
    \begin{equation}
        \begin{split}
        & \textstyle\sum_{s=1}^{W_{m,i}+S_{m,i}} A_m(t'_{m,i-1}+s) = \sum_{s=1}^{W_{m,i}+S_{m,i}} (S_{m,i-1}+s)\\
        & \textstyle\hspace{0.3cm}= S_{m,i-1}(W_{m,i}+S_{m,i}) + \frac{(W_{m,i}+S_{m,i})(W_{m,i}+S_{m,i}+1)}{2}.
        \end{split}
    \end{equation}
    Using the renewal reward theorem, we have
    \begin{equation}
        \begin{split} 
        \bar{A}_m & \textstyle= \frac{\mathbb{E}\left[S_{m,i-1}(W_{m,i}+S_{m,i}) + \frac{(W_{m,i}+S_{m,i})(W_{m,i}+S_{m,i}+1)}{2}\right]}{\mathbb{E}[W_{m,i}+S_{m,i}]} \\
        & \textstyle= \frac{\mathbb{E}[(W_{m,i}+S_{m,i})^2]}{2\mathbb{E}[W_{m,i}+S_{m,i}]} +\frac{\mathbb{E}[S_{m,i-1}(W_{m,i}+S_{m,i})]}{\mathbb{E}[W_{m,i}+S_{m,i}]} +  \frac{1}{2}.
        \end{split}
    \end{equation}

Further, since the average peak age $A^{\text{peak}}_m(i)$ for the $i^{th}$ update at cell $\omega_m$ is given by 
\begin{align*}
    A^{\text{peak}}_m(i) &= (t'_{m,i} - t_{m,i}) + (t_{m,i} - t'_{m,i-1}) + (t'_{m,i-1} - t_{m,i-1}) \\
    &= S_{m,i} + W_{m,i} + S_{m,i-1},
\end{align*}
we have $\bar{A}^{\text{peak}}_m = 2\mathbb{E}[S_{m,i}] + \mathbb{E}[W_{m,i}]$.

\subsection{Proof of Lemma~\ref{lemma:lower_bound}} \label{appendix:lower_bound}

    From Jensen's inequality, we have
    \begin{equation*}
        \mathbb{E}[(W_{m,i}+S_{m,i})^2] \ge (\mathbb{E}[W_{m,i}+S_{m,i}])^2.
    \end{equation*}
    Thus, we have
    \begin{equation*}
         \bar{A}_m \ge \frac{\mathbb{E}[(W_{m,i}+S_{m,i})^2]}{2\mathbb{E}[W_{m,i}+S_{m,i}]} + \frac{1}{2} \ge \frac{\mathbb{E}[W_{m,i}+S_{m,i}]}{2} + \frac{1}{2}.
    \end{equation*}

    Let $u_{m,k}(i)\in \{0,1\}$, where $u_{m,k}(i)=1$ if node $k$ updates in cycle $i$ of cell $\omega_m$. Then, $l_k u_{m,k}(i)$ is the number of fresh packets actually transmitted in cycle $i$. We can rewrite the long-term rate of updates using $\gamma_{m,k} = \mathbb{E}[u_{m,k}(i)]$ as
    

    \begin{equation}
         \sum_{k\in\mathcal{K}} \gamma_{m,k} = \frac{1}{\mathbb{E}[W_{m,i}]+\mathbb{E}[S_{m,i}]}.
    \end{equation}

      From (\ref{eq:act_freq}) and the communication constraint, we have 
     \begin{equation}
         \sum_{k\in \mathcal{K}_m}\frac{l_k\gamma_{m,k}}{p_kf_{m,k}} \le\sum_{k\in \mathcal{K}}\frac{l_k\nu_{k}}{p_k} \le 1,\nonumber
    \end{equation}
    and thus
    \begin{equation}
         \sum_{\omega_m\in V}\sum_{k\in \mathcal{K}_m}\frac{l_k \gamma_{m,k}}{p_kf_{m,k}} \le |V|. 
    \end{equation}
    
    Then, we can obtain a lower bound by solving
    \begin{equation}
        \begin{split}
            LB & = \frac{1}{|V|} \sum_{\omega_m \in V} \alpha_m \left(\frac{1}{2\sum_{k\in\mathcal{K}_m} \gamma_{m,k}} + \frac{1}{2}\right) \\
            &s.t.  \sum_{\omega_m \in V} \sum_{k \in \mathcal{K}_m} \frac{l_k \gamma_{m,k}}{p_k f_{m,k}}  \le |V|. \label{eq:age_lower_bound}
        \end{split}    
    \end{equation}
    
    From Cauchy-Schwarz inequality, Jensen's inequality and the communications constraint in (\ref{eq:age_lower_bound}), we have
    \begin{equation*}
        \begin{split}
            &\frac{1}{2} \left(\sum_{\omega_m \in V} \sqrt{\frac{\alpha_m \sum_{k\in \mathcal{K}_m}l_k\gamma_{m,k}}{ (\sum_{k\in \mathcal{K}_m}p_k f_{m,k})(\sum_{k\in \mathcal{K}_m} \gamma_{m,k})}}\right)^2 \\
            &\hspace{0.6cm} \le \left(\sum_{\omega_m\in V} \frac{\alpha_m  }{2 \sum_{k \in \mathcal{K}_m} \gamma_{m,k}} \right)\left( \sum_{\omega_m\in V} \frac{\sum_{k \in \mathcal{K}_m} l_k \gamma_{m,k}}{\sum_{k\in\mathcal{K}_m} p_k f_{m,k}}\right) \\
            &\hspace{0.6cm} \le \left(\sum_{\omega_m\in V} \frac{\alpha_m }{2\sum_{k\in \mathcal{K}_m} \gamma_{m,k}} \right)\left( \sum_{\omega_m} \sum_{k\in\mathcal{K}_m} \frac{l_k \gamma_{m,k}}{ p_kf_{m,k}}\right) \\
            &\hspace{0.6cm} \le |V| \left(\sum_{\omega_m\in V} \frac{\alpha_m }{2 \sum_{k \in \mathcal{K}_m} \gamma_{m,k}} \right).
        \end{split}
    \end{equation*}
    Hence, we have
    \begin{align*}
        &\mathbb{E}[J_\pi] \ge \frac{1}{2|V|}\sum_{\omega_m \in V} \alpha_m\\
        &+ \frac{1}{2|V|^2} \left(\sum_{\omega_m \in V} \sqrt{\frac{\alpha_m \sum_{k\in \mathcal{K}_m}l_k\gamma_{k,m}}{ (\sum_{k\in \mathcal{K}_m}p_kf_{k,m})(\sum_{k\in \mathcal{K}_m} \gamma_{k,m})}}\right)^2. \nonumber
    \end{align*}

\subsection{Proof of Proposition~\ref{proposition:without_sat}} \label{appendix:without_sat}

We first derive the expressions for the service time $S_m$. An update for cell $\omega_m$ is completed when a node covering the cell transmits all $l_k$ packets successfully. Since the first packet is already accounted for in $W_m$, the additional service time is due to the remaining $l_k-1$ packets. For node $k$, the additional service time is the sum of $l_k-1$ independent transmission attempts, each succeeding with probability $p_k$. Thus, the additional service time $\tilde{S}_k$ for node $k$ follows a negative binomial distribution with parameters $r = l_k-1$ and $p = p_k$; that is, $\tilde{S}_k \sim \text{NB}(l_k-1, p_k)$. Hence, for node $k$ we have
\begin{equation}
 \textstyle\mathbb{E}[\tilde{S}_k] = \frac{l_k-1}{p_k}, \quad \mathbb{E}[\tilde{S}_k^2] = \frac{(l_k-1)(l_k-p_k)}{p_k^2}.
\end{equation}

Let $B_{m} = \sum_{k \in \mathcal{K}_*} \mu_k p_k f_{m,k}$
denote the probability that a scheduled node can update cell $\omega_m$. Thus, conditioned on node $k$ being the one that updates cell $\omega_m$, the expected additional service time is
\begin{equation}
\textstyle\mathbb{E}[S_m \mid \text{update by node } k] = \frac{l_k-1}{p_k}.
\end{equation}
When a node is scheduled, the probability that node $k$ is the one selected (and is capable of updating cell $\omega_m$ is given by
\begin{equation}
\textstyle\mathbb{P}(\text{node } k \text{ updates } \omega_m) = \frac{\mu_k p_k f_{m,k}}{B_{m}}.
\end{equation}
By applying the law of total expectation over all nodes $k \in \mathcal{K}_*$, we obtain
\begin{align}
\mathbb{E}[S_m] &\textstyle= \sum_{k \in \mathcal{K}_*} \mathbb{P}(\text{node } k \text{ updates } \omega_m)\mathbb{E}[S_m \mid \text{update by node } k] \nonumber\\
&=  \textstyle\frac{1}{B_{m}}\sum_{k \in \mathcal{K}_*} \mu_k f_{m,k}(l_k-1).
\end{align}

Similarly, for the second moment, by conditioning on node $k$ we have
\begin{align}
\mathbb{E}[S_m^2] & \textstyle= \sum_{k \in \mathcal{K}_*} \mathbb{P}(\text{node } k \text{ updates } \omega_m)\mathbb{E}[S_k^2] \nonumber\\
&=  \textstyle\frac{1}{B_{m}}\sum_{k \in \mathcal{K}_*} \frac{\mu_k\,f_{m,k}(l_k-1)(l_k-p_k)}{p_k}.
\end{align}

Next, we analyze the waiting time $W_m$ for cell $\omega_m$. Suppose that cell $\omega_m$ has just been updated, and the scheduler initiates the next update process. At each time slot, a node $k \in \mathcal{K}_*$ is selected with probability $\mu_k$. The ability of node $k$ to update cell $\omega_m$ depends on $f_{m,k}(t)$, which is $1$ if the node can cover the cell at time $t$ and $0$ otherwise. This process can be viewed as a first-passage problem, where we seek the expected time until a node that is capable of updating cell $\omega_m$ is both scheduled and successfully transmits its update. We consider three cases:

\noindent {\bf Case 1 (Coverage):}  
If a scheduled node $k$ covers cell $\omega_m$ (i.e., $f_{m,k}(t)=1$), then the waiting time $W_m$ is determined by whether the first transmitted packet is successful:
\begin{equation}
 \textstyle W_m =
\begin{cases}
1, & \text{with probability } p_k, \\
1 + W_m, & \text{with probability } 1 - p_k.
\end{cases}
\end{equation}
Taking the conditional expectation yields
\begin{equation}
\mathbb{E}[W_m \mid f_{m,k}(t)=1] = p_k \cdot 1 + (1 - p_k)(1 + \mathbb{E}[W_m]).
\end{equation}

\noindent {\bf Case 2 (No Coverage):}  
If a scheduled node $k$ does not cover cell $\omega_m$ (i.e., $f_{m,k}(t)=0$), it uses the channel for its own update, thereby blocking cell $\omega_m$. Let $Y_k$ denote the blocking time due to node $k$. Then,
\begin{equation}
Y_k =
\begin{cases}
1, & \text{with probability } 1 - p_k, \\
1 + S_k, & \text{with probability } p_k,
\end{cases}
\end{equation}
where $S_k$ is the service time for node $k$'s update. Thus, the waiting time satisfies
\begin{equation}
W_m = Y_k + W_m.
\end{equation}
Taking expectations gives
\begin{equation}
\mathbb{E}[W_m \mid f_{m,k}(t)=0] = \mathbb{E}[Y_k] + \mathbb{E}[W_m] = l_k + \mathbb{E}[W_m],
\end{equation}
since $\mathbb{E}[Y_k] = (1-p_k)\cdot 1 + p_k\,(1+\mathbb{E}[S_k]) = l_k$.

\noindent {\bf Case 3 (Idle):} If no node is scheduled (which occurs with probability $1-\sum_{k\in \mathcal{K}_*} \mu_k$), then the channel remains idle for one slot and the system state is unchanged. Hence, we have $W_m = 1 + W_m$.

By the law of total expectation, we have
\begin{align}
\mathbb{E}[W_m] & \textstyle= \sum_{k\in\mathcal{K}_*} \mu_k f_{m,k} \mathbb{E}[W_m \mid f_{m,k}(t)=1] \nonumber \\
&\hspace{1cm} \textstyle+ \sum_{k\in\mathcal{K}_*} \mu_k (1-f_{m,k}) \mathbb{E}[W_m \mid f_{m,k}(t)=0] \\
&\hspace{1cm} \textstyle+ \left(1-\sum_{k\in\mathcal{K}_*}\mu_k \right)(1+\mathbb{E}[W_m]). \nonumber
\end{align}
Substituting the conditional expectations from above and rearranging terms leads to the expression for $\mathbb{E}[W_m]$ as
\begin{equation}
 \textstyle\mathbb{E}[W_m] = \frac{1}{B_{m}} \left(1+ \sum_{k\in\mathcal{K}_*} \mu_k (1-f_{m,k})(l_k-1)\right).
\end{equation}

A similar first-passage analysis is used to derive $\mathbb{E}[W_m^2]$. For the coverage case, we have
\begin{align}
\mathbb{E}[W_m^2 \mid f_{m,k}(t)=1] &= p_k \cdot 1^2 + (1-p_k) \mathbb{E}[(1+W_m)^2]\nonumber\\
&= 1 + (1-p_k) (2\,\mathbb{E}[W_m] + \mathbb{E}[W_m^2]).
\end{align}
For the no-coverage case,
\begin{equation}
\mathbb{E}[W_m^2 \mid f_{m,k}(t)=0] = \mathbb{E}[Y_k^2] + 2\,\mathbb{E}[Y_k]\,\mathbb{E}[W_m] + \mathbb{E}[W_m^2],
\end{equation}
where
\begin{equation}
 \textstyle\mathbb{E}[Y_k^2] = (1-p_k)\cdot 1^2 + p_k \mathbb{E}[(1+S_k)^2] = 2 l_k - 1 + \frac{(l_k-1)(l_k-p_k)}{p_k}. \nonumber
\end{equation}
Collecting the terms and rearranging yields the expression for $\mathbb{E}[W_m^2]$ as
\begin{align}
\mathbb{E}[W_m^2] & \textstyle= \frac{1}{B_{m}}\Bigl[\sum_{k \in \mathcal{K}_*} \mu_k f_{m,k}\Bigl[1 + 2(1-p_k)\mathbb{E}[W_m]\Bigr] \nonumber \\
&\hspace{1cm} \textstyle+ \sum_{k \in \mathcal{K}_*} \mu_k(1-f_{m,k})\Bigl[\mathbb{E}[Y_k^2] + 2l_k\mathbb{E}[W_m]\Bigr] \\ 
&\hspace{1cm} \textstyle+ \left(1-\sum_{k\in\mathcal{K}_*}\mu_k \right)(1+2\mathbb{E}[W_m]) \Bigr]. \nonumber
\end{align}

\subsection{Proof of Proposition~\ref{proposition:with_sat}} \label{appendix:with_sat}

    If a satellite is scheduled, it can cover all cell $\omega_m$. However, satellite transmissions may fail due to unavailability during on-going transmissions. That is,  given that the satellite successfully transmits its first packet, it is required that the time for completing the remaining $l_{\text{sat}}-1$ packets is less than (or equal to) the satellite's remaining available time. Let $\gamma$ denote the successful probability of a satellite update. 

    Recall that $T_A$ is the geometric random variable denoting the duration the satellite stays in the available state (A), and $T_U$ similarly represents the duration it remains in the unavailable state (U). Then, the steady-state probability that the satellite is in state A is given by $\pi_A = \frac{\lambda_A}{\lambda_A+\lambda_U}$.
    Further, recall that $S_{sat}$ is the total number of transmissions needed for $l_{\text{sat}}-1$ successful packets. The event of successful completion is $\{S_{sat} \le T_A\}$. Note that if $l_{sat} = 1$, then $S_{sat} = 0$ and thus $\mathbb{P}(0\le T_A)=1$. So, we consider $l_{sat}\ge 2$. By the memoryless property of $T_A$, the remaining time in state A (after any partial usage) is still geometric with the same parameter. Hence, we have
    \begin{align}
        \gamma & \textstyle= \mathbb{P}(S_{sat} \le T_A) = \sum_{r=l_{sat}-1}^\infty \mathbb{P}(S_{sat}=r)\mathbb{P}(T_A \ge r),
    \end{align}
    where $\mathbb{P}(S_{sat}=r) = \binom{r-1}{l_{sat}-2} p_{sat}^{l_{sat}-1} (1-p_{sat})^{r-(l_{sat}-1)}$
    for $r\ge l_{sat}-1$, and $\mathbb{P}(T_A \ge r)= (1-\lambda_A)^{r-1}$.
    Rearranging this, we can rewrite as
    \begin{align}
        \gamma & \textstyle= p_{sat}^{l_{sat}-1} (1-\lambda_A)^{l_{sat}-2 }\sum_{m=0}^\infty \binom{l_{sat}-2+m}{l_{sat}-2} [(1-p_{sat})(1-\lambda_A)]^m \nonumber \\
        & \textstyle= p_{sat}^{l_{sat}-1} (1-\lambda_A)^{l_{sat}-2 } \frac{1}{(1-(1-p_{sat})(1-\lambda_A))^{l_{sat}-1}} \nonumber \\
        & \textstyle= \frac{1}{1-\lambda_A}\left(\frac{p_{sat}(1-\lambda_A)}{p_{sat}+\lambda_A -p_{sat}\lambda_A}\right)^{l_{sat}-1}. \label{eq:gamma}
    \end{align}

    We now analyze the service time $S_m$ for cell $\omega_m$. For IoT and UAV updates, we have already analyzed the service time in the absence of satellite access. Hence, we analyze the satellite service time $S_{sat}$. Since an update can fail if the satellite becomes unavailable during the transmission, we are interested in the conditional mean:
    \begin{equation}
         \textstyle\mathbb{E}[S_{sat}\mid S_{sat} \le T_A] = \frac{\mathbb{E}[S_{sat} \cdot \mathbb{I}\{S_{sat}\le T_A\}]}{\mathbb{P}(S_{sat} \le T_A)},
    \end{equation}
    where the denominator is given in (\ref{eq:gamma}). To compute the numerator, we use the probability generating function $G_{S_{sat}}(z)$ of $S_{sat}$, which is given by
    \begin{equation}
         \textstyle G_{S_{sat}}(z) = \left(\frac{p_{sat} z}{1-(1-p_{sat})z}\right)^{l_{sat}-1}.
    \end{equation}
    Then, we have $\mathbb{E}[S_{sat} \cdot z^{S_{sat}}] = z\cdot G'_{S_{sat}}(z)$,
    where $ G'_{S_{sat}}(z)$
    \begin{align}
        & \textstyle= (l_{sat}-1) \left(\frac{p_{sat} z}{1-(1-p_{sat})z}\right)^{l_{sat}-2}  \frac{p_{sat}[1-(1-p_{sat})z] + p_{sat}z(1-p_{sat})}{[1-(1-p_{sat})z]^2}. \nonumber
    \end{align}

    We evaluate this at at $z=1-\lambda_A$, which corresponds to the probability $\mathbb{P}(T_A \ge r) = (1-\lambda_A)^{r-1} = z^{r-1}$. This allows us to rewrite the conditional expectation $\mathbb{E}[S_{sat}\mid S_{sat} \le T_A]$ as $\mathbb{E}[S_{sat} \cdot z^{S_{sat}-1}] = \frac{1}{z} \mathbb{E}[S_{sat} \cdot z^{S_{sat}}]$. With $z=1-\lambda_A$, we obtain
    \begin{align}
         \textstyle\mathbb{E}[S_{sat} \cdot z^{S_{sat}}] = \frac{(l_{sat}-1)p_{sat}(1-\lambda_A)}{(p_{sat}(1-\lambda_A)+\lambda_A)^2} \left(\frac{p_{sat}(1-\lambda_A)}{p_{sat}(1-\lambda_A)+\lambda_A}\right)^{l_{sat}-2}. 
    \end{align}
    Combining this with (\ref{eq:gamma}), we have
    \begin{equation}
         \textstyle\mathbb{E}[S_{sat}\mid S_{sat} \le T_A] = \frac{l_{sat}-1}{p_{sat} + \lambda_A -p_{sat}\lambda_A}.
    \end{equation}
    
    Let $C_{m} = \sum_{k\in\mathcal{K}_*} \mu_k p_k f_{m,k} + \pi_A \mu_{sat} p_{sat} \gamma$.
    Taking into account potential satellite update failures, and following the same reasoning as in the no-satellite case, we have
    \begin{equation}
         \textstyle \mathbb{E}[S_m] = \frac{1}{C_m} \left(\sum_{k\in\mathcal{K}_*} \mu_k f_{m,k}(l_k-1) + \frac{\pi_A\mu_{sat} p_{sat}\gamma (l_{sat}-1)}{p_{sat} + \lambda_A -p_{sat}\lambda_A} \right).
    \end{equation}

    Similarly, we can compute $\mathbb{E}[S_{sat}^2 \mid S_{sat} \le T_A]$ using the probability generating function $G_{S_{sat}}(z)$, denoted simply as $G(z)$ for brevity. To reduce notational complexity, we omit the subscript ‘$sat$’ in $p_{sat}$ and $l_{sat}$ and let $r = l - 1$.
    \begin{align}
        \mathbb{E}[S^2_{sat}\mid S_{sat} \le T_A] & \textstyle= \frac{\mathbb{E}[S^2_{sat}\mathbb{I}\{ S_{sat} \le T_A\}]}{\mathbb{P}\{ S_{sat} \le T_A\}}  = \frac{zG'(z) +z^2 G''(z)}{G(z)}. \label{eq:ES2_pre} 
    \end{align}
    Let  $a(z) = \frac{pz}{1-(1-p)z}$ and $b(z) = 1-(1-p)z$.
    Then, we have
    \begin{equation}
         \textstyle G(z) = a(z)^r, \quad a'(z) = \frac{p}{b(z)^2}, \quad a''(z) = \frac{2p(1-p)}{b(z)^3}.
    \end{equation}
    The derivatives of $G(z)$ can be expressed as
    \begin{align}
        G'(z) &= ra(z)^{r-1} a'(z), \\
        G''(z)&= r(r-1)a(z)^{r-2}(a'(z))^2 + ra(z)^{r-1} a''(z). 
    \end{align}
    Substituting into (\ref{eq:ES2_pre}), we obtain
    \begin{align}
        \mathbb{E}[S^2_{sat}\mid S_{sat} \le T_A] & \textstyle= \frac{rza'(z)}{a(z)} + \frac{r(r-1)z^2(a'(z))^2}{a(z)^2} + \frac{rz^2a''(z)}{a(z)}\nonumber \\
        &\hspace{-1cm} \textstyle= \frac{r}{p+\lambda_A-p\lambda_A} + \frac{r(r-1+2(1-\lambda_A)(1-p))}{(p+\lambda_A -p\lambda_A)^2}. 
    \end{align}
    Rewriting this in terms of the original parameters, we have
    \begin{align}
        \mathbb{E}[S^2_{sat}\mid S_{sat} \le T_A] & \textstyle= \frac{(l_{sat}-1)(l_{sat}-(p_{sat} + \lambda_A -p_{sat}\lambda_A)))}{(p_{sat} + \lambda_A -p_{sat}\lambda_A)^2}, \nonumber
    \end{align}
    and thus we have
    \begin{align}
        \mathbb{E}[S_m^2] & \textstyle= \frac{1}{C_m} \big( \sum_{k \in \mathcal{K}_*} \frac{\mu_k f_{m,k}(l_k - 1)(l_k -p_k)}{p_k}  \\
        &\hspace{0.2cm} \textstyle+ \frac{\pi_A\mu_{sat} p_{sat}\gamma (l_{sat}-1)(l_{sat}-(p_{sat} + \lambda_A -p_{sat}\lambda_A)))}{(p_{sat} + \lambda_A -p_{sat}\lambda_A)^2}\big). \nonumber
    \end{align}

    We next analyze the waiting time $W_m$ for cell $\omega_m$. When the satellite is in the unavailable state (U), it cannot participate in updates. In this case, the system behavior is identical to that of the scenario without satellite support. In contrast, during the available state (A), the satellite becomes part of the update process. Hence, in addition to the three cases in the no-satellite case, we consider the following case.
    
    \noindent \textbf{Case 4 (Satellite):} If a satellite is scheduled, it can cover all cell $\omega_m$. However, satellite transmissions may fail due to unavailability during the transmission window. Let $X_{w}$ denote the wasted time (i.e., the time until the satellite becomes unavailable after the first packet transmission of the satellite), and note that $T_u$ denotes the satellite unavailable time with $\mathbb{E}[T_u] = \frac{1}{\lambda_U}$. Then, we have
    \begin{equation}
     \textstyle W_m = 
    \begin{cases}
        1, & \text{w.p. } p_{sat} \gamma, \\
        1 + W_m, & \text{w.p. } 1 - p_{sat}, \\ 
        1+X_{w} + T_u + W_m, &\text{w.p. } p_{sat}(1-\gamma).        
    \end{cases}
    \end{equation}
    Combining with the three cases and rearranging it, we have
    \begin{align}
        \mathbb{E}[W_m] & \textstyle= \frac{1}{C_{m}} \big(1+ \sum_{k\in\mathcal{K}_*} \mu_k (1-f_{m,k})(l_k-1) \\
        &\hspace{1cm} \textstyle+ \pi_A\mu_{sat}p_{sat}(1-\gamma)(\mathbb{E}[X_{waste}]+\frac{1}{\lambda_U})\big), \nonumber
    \end{align}
    where
    \begin{align}
        \mathbb{E}[X_{w}] & \textstyle= \mathbb{E}[T_A \mid T_A < S_{sat}]  = \frac{\sum_{k=1}^{\infty} k\mathbb{P}(T_A=k)\mathbb{P}(S_{sat}>k)} {\sum_{k=1}^{\infty} \mathbb{P}(T_A=k)\mathbb{P}(S_{sat}>k)} \nonumber\\
        &\hspace{-1.2cm} \textstyle= \frac{\sum_{k=1}^\infty k\lambda_A(1-\lambda_A)^{k-1}(1-I_{p_{sat}}(l_{sat}-1,k-l_{sat}+2))}{\sum_{k=1}^\infty \lambda_A(1-\lambda_A)^{k-1}(1-I_{p_{sat}}(l_{sat}-1,k-l_{sat}+2))}, 
    \end{align}
    where $I_p(a,b)$ is the regularized incomplete Beta function.

    Similarly, we can obtain
    \begin{align}
        \mathbb{E}[W_m^2] &\textstyle= \frac{1}{C_{m}}\Bigl[\sum_{k \in \mathcal{K}_*} \mu_k f_{m,k}\Bigl[1 + 2(1-p_k)\mathbb{E}[W_m]\Bigr] \nonumber \\
        &\textstyle \hspace{-1cm}+ \sum_{k \in \mathcal{K}_*} \mu_k(1-f_{m,k})\Bigl[\mathbb{E}[Y_k^2] + 2l_k\mathbb{E}[W_m]\Bigr]\nonumber \\ 
        &\textstyle \hspace{-1cm}+ \left(1-\sum_{k\in\mathcal{K}_*}\mu_k - \pi_A \mu_{sat} (1+p_{sat}) \right)(1+2\mathbb{E}[W_m]) \\
        &\textstyle \hspace{-1cm}+  \pi_A \mu_{sat} p_{sat}\Big(\gamma + (1-\gamma)\big(\mathbb{E}[X_{w}^2] + 2\mathbb{E}[X_{w}]\mathbb{E}[W_m]\nonumber\\
        &\textstyle\hspace{-1cm}+  2\mathbb{E}[X_{w}]+2\mathbb{E}[W_m]+ \frac{2}{\lambda_U^2}(1+\lambda_U(1+\mathbb{E}[X_{w}]+ \mathbb{E}[W_m])) +1  \big) \Big)\Bigr], \nonumber
    \end{align}
    where
    \begin{equation}
         \textstyle\mathbb{E}[X^2_{w}] = \frac{\sum_{k=1}^\infty k^2\lambda_A(1-\lambda_A)^{k-1}(1-I_{p_{sat}}(l_{sat}-1,k-l_{sat}+2))}{\sum_{k=1}^\infty \lambda_A(1-\lambda_A)^{k-1}(1-I_{p_{sat}}(l_{sat}-1,k-l_{sat}+2))}.
    \end{equation}




\end{document}